\documentclass[twocolumn,tighten]{aastex63}

\usepackage{color}

\def\um{{\sc UniverseMachine}}
\def\omegachired{\omega_{\chi,r}}
\def\omegachiblue{\omega_{\chi,b}}
\def\lgrp{L_{\rm grp}}
\def\psat{P_{\rm sat}}
\def\chitot{\chi^2_{\rm tot}}
\def\wp{w_p(r_p)}
\def\wchi{w_\chi}
\def\wchic{w_{\chi,c}}
\def\wchir{w_{\chi,r}}
\def\wchib{w_{\chi,b}}
\def\wcen{w_{\rm cen}}
\def\wcenc{w_{\rm cen,c}}
\def\wcenr{w_{\rm cen,r}}
\def\wcenb{w_{\rm cen,b}}
\def\sfr{\rm SFR}
\def\ssfr{\rm sSFR}
\def\sigv{\sigma_v}
\def\mr{M_r-5\log h}
\def\lgal{L_{\rm gal}}
\def\lsolhh{h^{-2}{\rm L}_\odot}
\def\mh{M_h}
\def\fsat{f_{\rm sat}}
\def\msol{$M$_\odot}
\def\hmsol{h^{-1}{\rm M}_\odot}
\def\mgal{M_\ast}
\def\mhalo{M_h}
\def\slogl{\sigma_{\log L}}
\def\slogm{\sigma_{\log M_\ast}}
\def\lsat{L_{\rm sat}}
\def\lsatred{L_{\rm sat}^{\rm red}}
\def\lsatb{L_{\rm sat}^{\rm blue}}
\def\lsatr{L_{\rm sat}^{\rm red}}
\def\lsatbar{\bar{L}_{\rm sat}}
\def\lsatblue{L_{\rm sat}^{\rm blue}}
\def\hkpc{h^{-1}{\rm kpc}}
\def\hmpc{h^{-1}{\rm Mpc}}
\def\mpeak{M_{\rm peak}}
\def\mocka{{\sc MockA}}
\def\mockb{{\sc MockB}}
\def\mockum{{\sc MockUM}}
\def\bsat{B_{\rm sat}}
\def\bsatc{B_{\rm sat,c}}
\def\bsatred{B_{\rm sat,r}}
\def\bsatblue{B_{\rm sat,b}}

\def\ltotw{{L}_{\rm tot}}
\def\lsatchibar{L_{\rm sat}(\chi|\lgal)/\lsat(\lgal)}

\submitjournal{ApJ}

\shorttitle{Self-Calibrating Group Finder}
\shortauthors{Tinker}

\begin{document}

\title{A Self-Calibrating Halo-Based Galaxy Group Finder: Algorithm and Tests}

\correspondingauthor{Jeremy L. Tinker}
\email{jeremy.tinker@nyu.edu}

\author[0000-0003-3578-6149]{Jeremy L. Tinker}
\affiliation{Center for Cosmology and Particle Physics, Department of
  Physics, New York University, New York, USA, 10003}

\begin{abstract}

  We describe an extension of the halo-based galaxy group-finding
  algorithm that adds freedom to the algorithm to more accurately
  determine which galaxies are central and which are satellites and to
  provide unbiased estimates of halo masses. We focus on determination
  of the galaxy-halo relations for star-forming and quiescent
  galaxies. The added freedom in the group-finding algorithm is
  self-calibrated using observations of color-dependent galaxy
  clustering, as well as measurements of the total satellite
  luminosity in deep imaging data around stacked samples of
  spectroscopic central galaxies, $\lsat$. We test this approach on a
  series of mocks that vary the galaxy-halo connection, including one
  mock constructed from UniverseMachine results.  Our self-calibrated
  algorithm shows marked improvement over previous methods in
  estimating the color-dependent satellite fraction of galaxies.  It
  reduces the error in $\log\mhalo$ for central galaxies by over a
  factor of two, to $\lesssim 0.2$ dex. Through the $\lsat$ data, it
  can quantify differences in the luminosity-to-halo mass relations
  for star-forming and quiescent galaxies, even for groups with only
  one spectroscopic member. Thus, whereas previous algorithms cannot
  constrain the scatter in $\lgal$ at fixed $\mhalo$, the
  self-calibration technique can provide a robust lower limit to this
  scatter.

\end{abstract}

\keywords{galaxies---groups: galaxies---halos}

\section{Introduction} \label{sec:intro}

Galaxy group catalogs are one of the key tools for advancing our
knowledge of galaxy formation and evolution (e.g.,
\citealt{huchra_geller:82, eke_etal:04, yang_etal:05,
  berlind_etal:06_catalog, robotham_etal:11,
  tinker_etal:11,gerke_etal:12}). To robustly interpret the spectrum
of galaxy properties, it is critical to understand which galaxies are
central in their dark matter halo, and which are satellites orbiting
within the halo potential. Armed with this ability to decompose the
galaxy population in the local universe into populations of central
and satellite galaxies, for example, \cite{peng_etal:12} found that the
correlations of galaxy color with environment were driven
by the rising fraction of satellite galaxies in high-density
regions, and that the colors of central galaxies were nearly
independent of environment (see also, \citealt{blanton_berlind:07,
  tinker_etal:08_voids, tinker_etal:11, tinker_etal:17_p1, elucid4}).

This central-satellite decomposition afforded by group catalogs has
had other notable successes in our understanding of galaxy
formation. This knowledge has yielded quenching timescales of
satellite galaxies (\citealt{weinmann_etal:10,
  wetzel_etal:13_groups2}), and central galaxies
(\citealt{hahn_etal:17_tq}). Groups catalogs have provided a
model-independent confirmation of the Halo Occupation Distribution
description of galaxy bias (HOD; \citealt{yang_etal:08}) as well as
observational constraints on the stellar-to-halo mass relation (SHMR;
see the recent review of the galaxy-halo connection by
\citealt{wechsler_tinker:18}). The information they provide is one of
the primary methods of investigating the galaxy-halo connection and
providing tests of galaxy formation models (e.g.,
\citealt{henriques_etal:17, delucia_etal:19}).

Although many different definitions of galaxy groups exist in the
literature, the definition that best facilitates these results is that
a group of galaxies is a set of galaxies that occupy a common dark
matter halo. In this vein, the true purpose of a galaxy group finder
is not finding groups of galaxies, but rather to find dark matter
halos and identify which galaxies reside are their centers. 

Despite the many successes of modern group finders, significant
problems in their efficacy hinder further progress. In this paper we
will address several of these issues to construct a new group finding
algorithm. All group findering algorithms contain within then free
parameters. Usually, these parameters are calibrated on mock galaxy
distributions. However, this means that the calibration of the group
finder is sensitive to the details of the mock construction and the
physical realism of the mock. Our solution is to add extra freedoms to
the group-finding algorithm that directly address their current
limitations, and then self-calibrate these parameters through
independent observations that indirectly probe the galaxy-halo
connection, such as clustering and cross-correlations with deep
imaging data.

\subsection{Problems in Current Group Finders}

Group finding algorithms typically fall into two classifications:
percolation-based finders and halo-based finders. Percolation-based
finders use the friends-of-friends algorithm to link together galaxies
in a group. This approach was first used on spectroscopic survey data
by \cite{huchra_geller:82}, and then further implemented on
large-scale galaxy redshift surveys in 2dFGRS (\citealt{eke_etal:04}),
SDSS (\citealt{berlind_etal:06_catalog}), and GAMA
(\citealt{robotham_etal:11}). Halo-based algorithms, developed by
\cite{yang_etal:05}, use our theoretical knowledge of propreties and
statistics of dark matter halos to infer membership probabilities of
central and satellite galaxies. We focus here on improving the
halo-based group finding algorithm, but we note that many of the
issues we will discuss are found in percolation-based methods as
well. \cite{campbell_etal:15} detailed many failures of both approaches,
especially in the context of understanding the galaxy-halo connection
for galaxy samples broken into star-forming and quiescent samples.

\vspace{0.2cm}
\noindent {\it 1. Impurities and incompleteness in satellite
  galaxies.} Given the fact that we don't know the true distance to
galaxies in a redshift survey, it is impossible to achieve perfect
purity and completeness when assessing which galaxies are satellites
within a sample. These errors are magnified when separately
quantifying the satellite fractions, $\fsat$, of red and blue
galaxies. Satellites have a higher quenched fraction relative to
central galaxies (\citealt{scranton:02, vdb_etal:03_early_late,
  zehavi_etal:05, zehavi_etal:11, weinmann_etal:06,
  wetzel_etal:12_groups1}). This is due to a number of physical
mechanisms that can heat, strip, or cut off the gas supply of the
satellite galaxy, none of which are available to central galaxies that
exist in field halos. Thus, when galaxies are misclassified, this
leads to an underestimate of the red satellite fraction and an
overestimate of the blue satellite fraction, even if the overall
$\fsat$ is accurately gauged. When trying to compare theoretical
models to observed group data, one solution to this problem is to
forward model the theory by constructing a mock galaxy sample and
passing it through the group finder, as done in
\cite{wetzel_etal:12_groups1,wetzel_etal:13_groups3} and
\cite{calderon_etal:18}. This is a cumbersome process and it not
always available if the theory is not applied to a simulation.

\vspace{0.2cm}
\noindent {\it 2. Estimating halo masses for sub-classes of central galaxies.}
Most current group finders of both varieties rely on the total
luminosity or stellar mass within a group to estimate the halo
mass. In this method, the assigned halo masses follow the
rank-ordering of the group luminosities. This becomes a problem when
the SHMR of one class of galaxies differs is a systematic way to that
of another class of galaxies.

Due to the fact that blue galaxies increase their stellar mass over
time, while red galaxies are largely stagnant after their arrival on the red
sequence
it is possible that red and blue subsamples have different
SHMRs. Further, the compilation of recent analyses of the red and blue
SHMRs in \cite{wechsler_tinker:18} shows that there is no consensus in
the community on this question.

The way this introduces errors into the group-finding process is as
follows: The central galaxy dominates the total luminosity and stellar
mass of the group for halo masses below $\sim 10^{14}$ $\msol$
(\citealt{yang_etal:03, tinker_etal:05, vale_ostriker:06,
  sheldon_etal:09_ml, leauthaud_etal:12_total, tinker_etal:19_lsat}),
even though it may have a number of satellite galaxies. Thus, in the
rank-ordering of total group luminosities, groups with red and blue
central galaxies of the same luminosity will be next to each other in
the rank-order, and thus be assigned the same dark matter halo mass,
regardless of the physical reality. As a consequence, group finders
can accrue errors of the host halo masses of nearly an order of
magnitude (\citealt{campbell_etal:15}).

\vspace{0.2cm}
\noindent {\it 3. No inclusion of the scatter between halos and
  central galaxies.} The scatter in the galaxy-halo connection is
generally parameterized as a lognormal scatter of $\mgal$ at fixed
$\mhalo$, with width $\slogm$ (or $\slogl$ for $\log\lgal$; see \S 4.3
in \citealt{wechsler_tinker:18} and references therein). This approach
is in reasonable agreement with the results of hydrodynamic
simulations and semi-analytic galaxy formation models
(\citealt{matthee_etal:17}). In the rank-ordering scheme of halo
assignment, the variation of $M_{\ast, \rm cen}$ at fixed $\mhalo$ is
perfectly correlated with total stellar mass in the satellite
galaxies, as $\mhalo$ is set by $M_{\ast, \rm tot}$. Thus, when the
number of members in a group is reasonably large, the value of
$\slogm$ in the group catalog is non-zero but typically smaller than
the true value (\citealt{reddick_etal:13, cao_etal:19}). However, when
the number of members is low, such that the total group luminosity is
dominated by the central galaxy, the true scatter is entirely
suppressed and central galaxies with the same luminosity will be
assigned the same halo mass. Thus the value of $\slogm$ in the catalog
approaches zero.

\subsection{Toward a group finder that solves these problem}

All galaxy grop finders have free parameters. For the \cite{yang_etal:05} algorithm,
the unknown quantity is the threshold probability above which a galaxy
can be considered a satellite of a larger halo, $\bsat$. For the
percolation algorithms of \cite{berlind_etal:06_catalog} and
\cite{robotham_etal:11}, it is the linking lengths for projected
separation and line-of-sight separation between neighboring
galaxies. In these cases, these unknowns are calibrated on mock galaxy
catalogs. Thus, the values that are used to analyze real data are
dependent on the details of the mock construction.

Although our knowledge of the galaxy-halo connection for luminosity or
stellar mass is robust, when extended to secondary galaxy properties,
such as color, star formation rate, morphology, or size, many
different choices can be made. For galaxy color and star-formation
rate, the mechanisms that influence galaxy growth and quenching are
not fully known. Whether halos play an active or passive role in these
processes is a current area of debate. For example, one may construct
toy models in which central galaxy quenching is only a function of
halo mass, or one in which quenching of these galaxies is entirely a
function of galaxy stellar mass. These models will yield different
SHMRs, even if they match observed quenched fractions. Additionally,
there are number of ways to parameterize the quenching of satellite
galaxies as well. Thus, the calibration of group finding free
parameters on mocks may influence the end results.

The alternative we explore here is to bring in external data
with which to self-calibrate the group finder's free parameters. The
primary output of a group finder is the statistical relationship
between halos and galaxies---the mean number of galaxies as a function
of halo mass and galaxy properties. If the group finder has done its
job effectively, then the halo occupation function (HOD) estimated by the
finder is representative of reality. A straightforward method of
testing this is to populate the halos in a simulation with the
occupation function derived by the group catalog, and compare the
resulting galaxy clustering to that measured in the actual data. If
the occupation function is wrong--for example, it underpredicts the
fraction of red galaxies that are satellites---then the clustering
will be different than the observations. This approach is related to
that taken by \cite{sinha_etal:18}, in which they use a group finder to
constrain the free parameters of the HOD. In our approach the HOD is
non-parametric, but the free parameters are within the group finder itself.

Clustering, however, provides limited influence on the mapping between
luminosity and halo mass for central galaxies in halos of
$\mhalo\lesssim 10^{13}$ $\msol$. Other information is required to
decide if a central galaxy lives in a halo of $\mhalo\sim 10^{11.5}$
$\msol$, or occupies a somewhat larger $10^{12}$ $\msol$ halo, neither
of which are likely to have any satellite galaxies in the
spectroscopic sample being analyzed.

To address this issue, we incorporate new results of the total
satellite luminosity around central galaxies, $\lsat$, from
\cite{tinker_etal:19_lsat} and \cite{alpaslan_tinker:19}. The $\lsat$
statistic cross-correlates central galaxies with faint imaging
catalogs, far fainter than the spectroscopic limit of the SDSS Main
Galaxy Sample (MGS; \citealt{strauss_etal:02}). These data supplement
the spectroscopic data, filling in the gaps of our ability to connect
galaxies to halos. \cite{alpaslan_tinker:19} demonstrated that, at
fixed stellar mass, galaxy secondary properties show strong
correlations with $\lsat$, implying that dark halos influence these
properties. This result agrees with recent weak lensing results of
KiDS+GAMA, in which directly measured correlations between halo mass
and galaxy secondary properties at fixed stellar mass
(\citealt{taylor_etal:20}).

One key result of \cite{alpaslan_tinker:19} is that stellar mass does
not contain the most information about $\lsat$, and, by extension,
$\mhalo$. In all cases, $r$-band absolute magnitude, $M_r$, of the
central correlated best with $\lsat$. Combined with persistent
theoretical uncertainties in estimating stellar masses (e.g.,
\citealt{lower_etal:20}), and the difficulty in of creating
volume-limited samples that are complete in stellar mass, these
results lead us to use $r$-band luminosity, $\lgal$, as our primary
quantity to map between galaxies and halos. In analogy with the SHMR,
our main quantities of interest are (1) the luminosity-to-halo mass
relation, LHMR, when broken into star-forming and quiscent samples,
and (2), the fraction of each sample that are satellites. For brevity,
we will use ``blue'' and ``red'' to refer to these types of galaxies.

In \S \ref{s.mocks}, we will describe the mocks that we use to test
out new algorithm. In \S \ref{s.methods}, we will review the current
halo-based group-finding algorithm, and present our extensions to
a self-calibrated model. In \S \ref{s.results}, we will show the
performance of our algorithm in recovering unbiased values of the
LHMRs, $\fsat$ values, and $\slogl$ values. We will summarize and
discuss prospects for application of this algorithm in \S
\ref{s.summary}. Most results here use the high resolution
Bolshoi-Planck cosmological N-body simulation
(\citealt{klypin_etal:16}).


\begin{figure}[ht!]
  \epsscale{1.2}
  \hspace{-0.4cm}
  \plotone{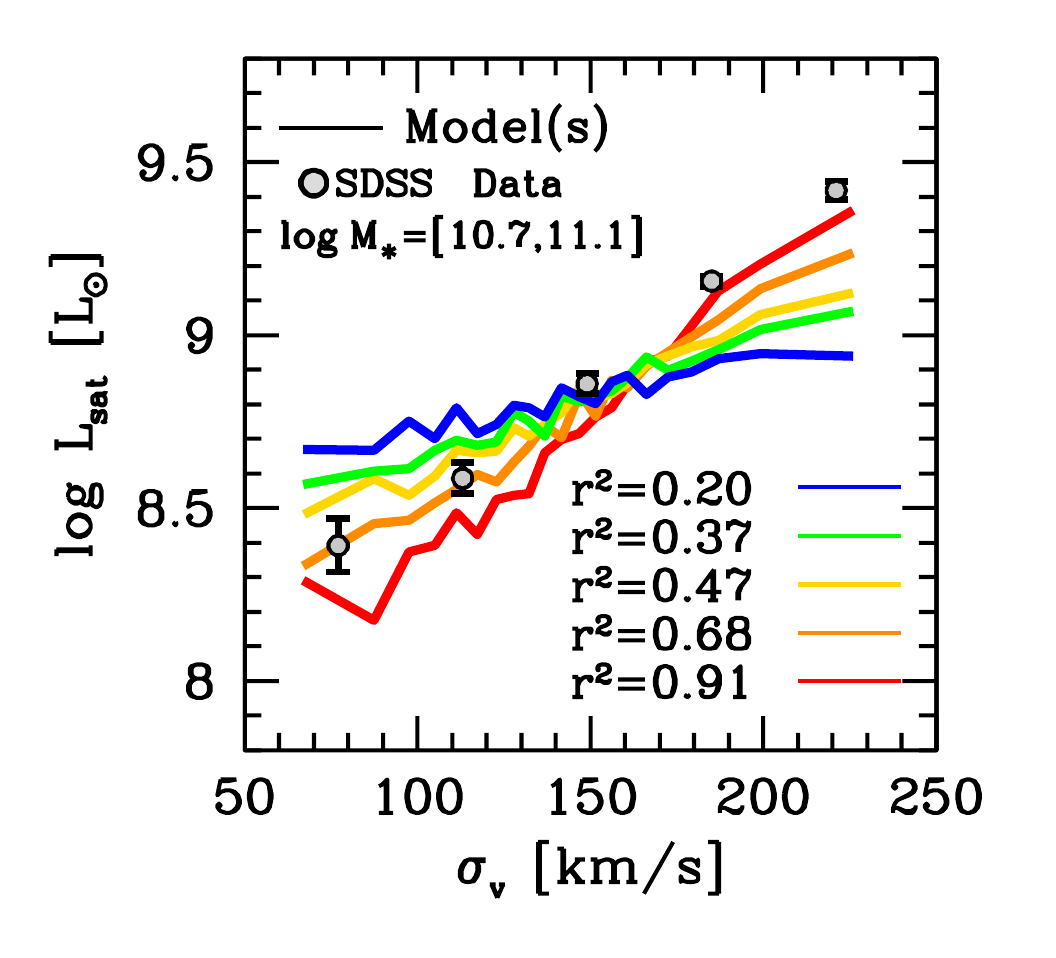}
  \vspace{-0.8cm}
  \caption{\label{f.sigv_corr} The correlation between $\lsat$ and
    $\sigv$ for a bin in galaxy stellar mass. Points with errors are
    taken from \cite{alpaslan_tinker:19}. The multi-colored curves
    show theoretical models that vary the correlation between $\mhalo$
    and $\sigma_v$ at fixed $\mgal$. In order to match the
    observations, the correlation needs to have $r^2\gtrsim 0.9$,
    implying that $\sigv$ contains significant information about
    $\mhalo$. These results are the motivation behind the inclusion of
    the hypothetical secondary galaxy property, $\chi$, in \mockb.  }
\end{figure}


\begin{figure*}[ht!]
\plotone{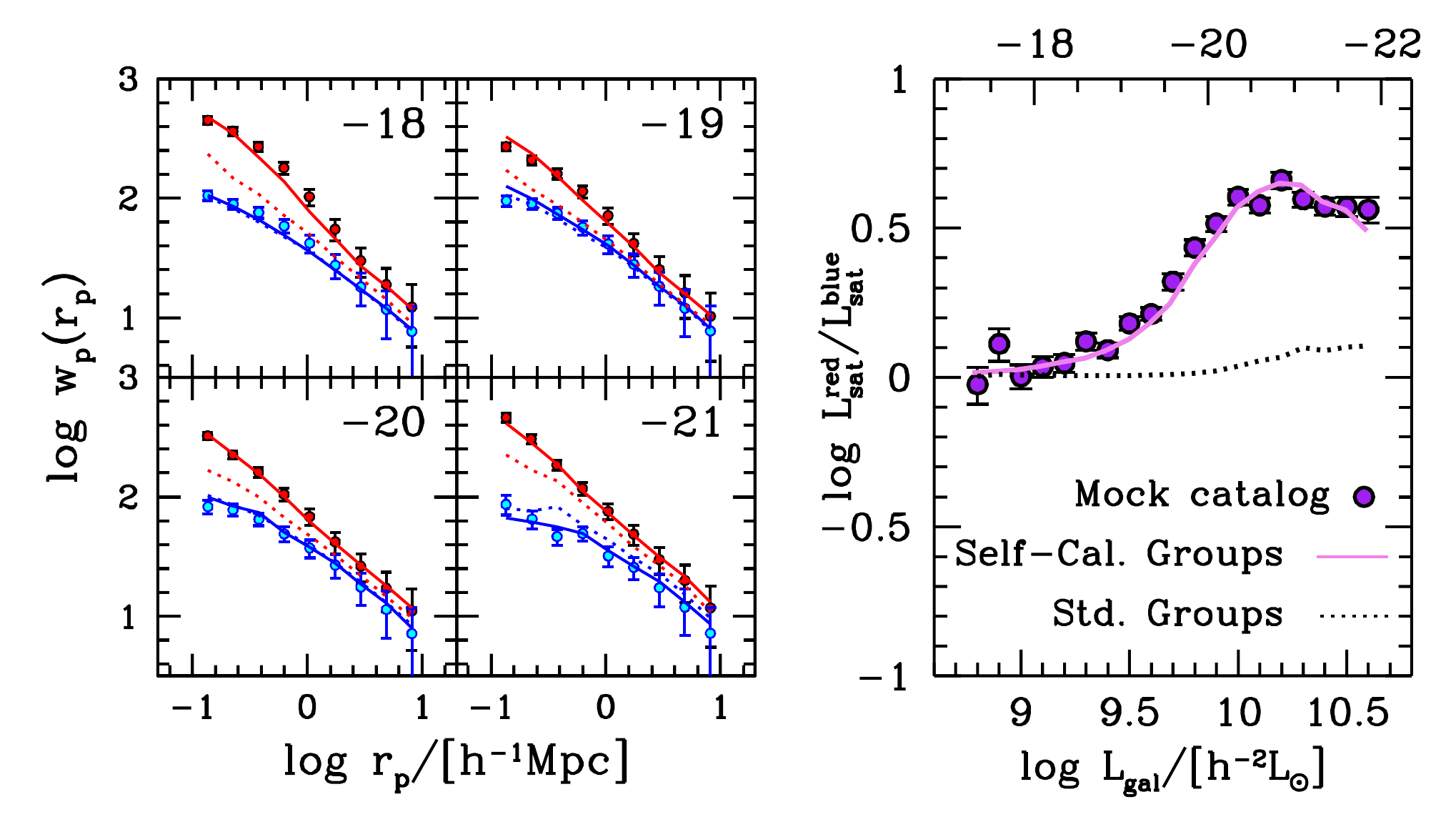}
\caption{ Observed properties of, and group finder fit to, \mocka. The
  quartet of panels on left-hand side of the figure show the projected
  correlation function in bins of $r$-band magnitude, for red and blue
  galaxies. To conserve space, we have not shown the $-17$ bin, but
  the results are similar to the $-18$ bin. The measurements from the
  mocks are indicated by the points with error bars; red and blue
  colors indicate red and blue galaxies. The solid curves are the
  best-fit self-calibrated group catalog. The dotted curves show the
  prediction from the standard group finder, without
  self-calibration. The large panel on the right-hand side shows the
  $\lsat$ ratio between red and blue galaxies at fixed $\lgal$. Points
  and curves are the same as the clustering
  panel.  \label{f.wp_lsat_mocka}}
\end{figure*}

\begin{figure*}[ht!]
\plotone{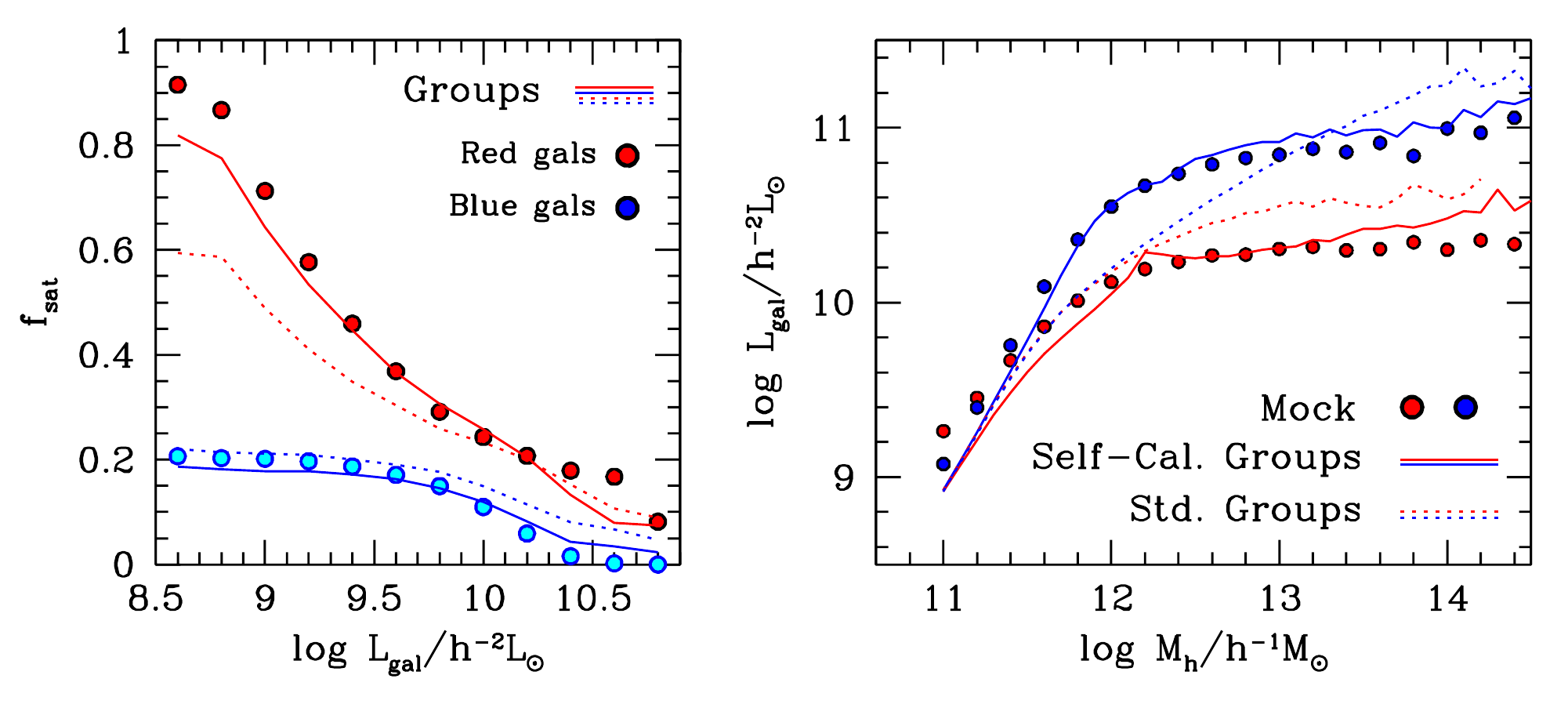}
\caption { {\it Left-Hand Panel:} Satellite fraction, $\fsat$, as a
  function of $\lgal$, for red and blue galaxies. Colored circles
  show values from \mocka. Dotted curves show the predictions from
  the standard group finder, which significantly underpredicts $\fsat$ for
  red galaxies. This is consistent with the clustering of red galaxies
  predicted by the standard finder in Figure \ref{f.wp_lsat_mocka}. The
  solid curves show the predictions from the self-calibrated group
  finder. {\it Right-Hand Panel:} The galaxy-halo connection for red
  and blue galaxies. Filled circles show the values from \mocka. Solid
  curves show the predictions from the self-calibrated group
  finder. As with $\fsat$, we stress that this is a prediction of the
  model. The prediction from the standard group finder, shown with the
  dotted curves, cannot distinguish the halos of red and blue central
  galaxies for $\lgal\lesssim 10^{10.4}$ $\lsolhh$. 
  \label{f.fsat_shmr_mocka}}
\end{figure*}

\begin{figure*}[ht!]
  \epsscale{1.2}
  \hspace{-0.5cm}
  \plotone{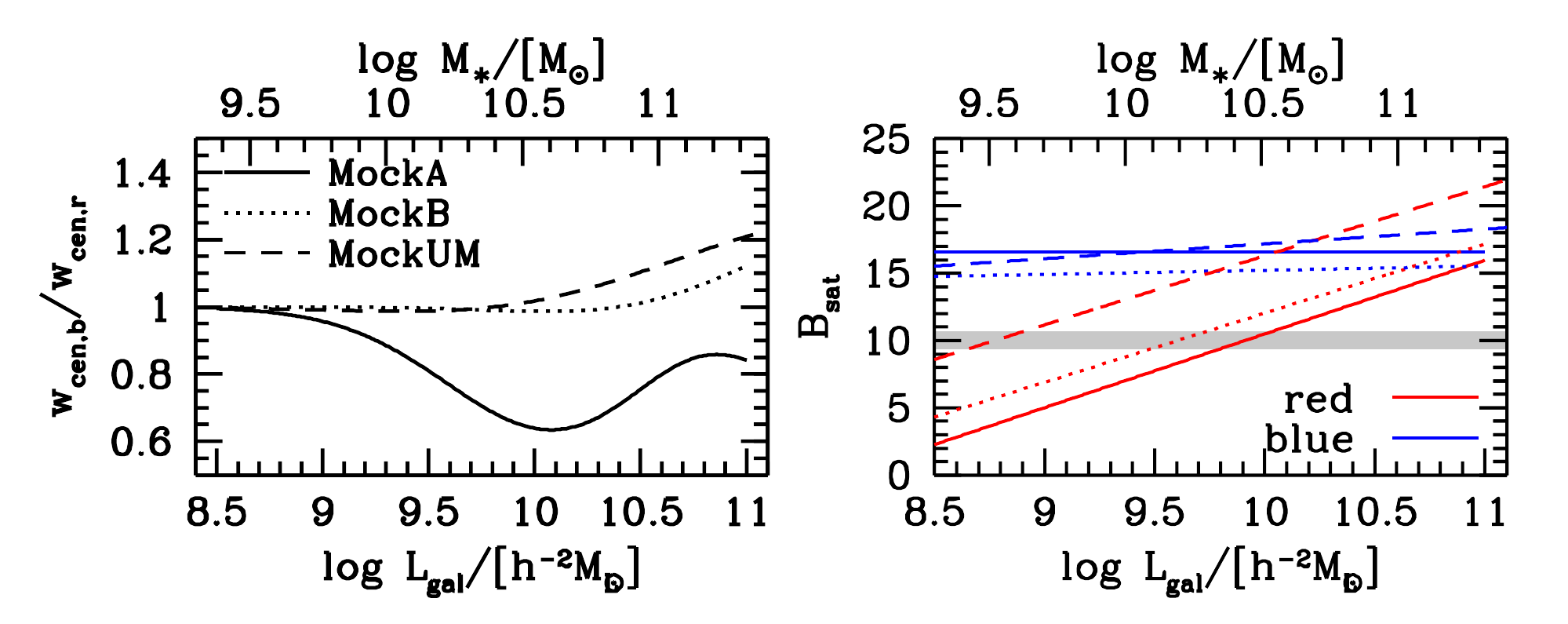}
  \vspace{-0.6cm}
  \caption{{\it Left Panel:} Relative weights assigned to blue and red
    central galaxies in the best-fit model for the three mocks.  These
    weights are used to determine the weighted total luminosity of the
    group, $\ltotw$.  The total luminosity is used in the
    rank-ordering of groups to assign halo mass via abundance matching
    via Eq.~(\ref{e.amtot}). The lower $x$-axis shows the luminosity
    of central galaxies in \mocka\ and \mockb, while the upper
    $x$-axis shows the stellar mass scale for \mockum.  {\it Right
      panel:} The best-fit model for the satellite
    threshold values for all three mocks. Line types, and $x$-axes are
    the same as the key in the left panel. The standard value of
    $\bsat=10$ is shown with the gray line.  \label{f.bestfit} }
\end{figure*}

\section{Mock Construction}
\label{s.mocks}

To test whether our self-calibration can work in myriad galaxy-halo
connections for red and blue subsamples of galaxies, we have compiled
a set of three mocks with different properties. In addition to two
mock we constructed, we have also applied our algorithm to public
catalogs from UniverseMachine (\citealt{behroozi_etal:19}). We use
more than one mock because we want to test different LHMRs, different
satellite abundances, and different input physics.

Both in-house mocks are constructed using the Bolshoi-Planck
simulation (\citealt{klypin_etal:16}). We use the abundance matching
technique to assign galaxy luminosity to dark matter halos. We use
$\mpeak$ as the halo quantity to match to $\lgal$. The simulation
volume is 250 $\hmpc$ per side, yielding a volume-limited angular
sample that extends to a redshift of $z=0.085$ with a footprint of
1/8$^{\rm th}$ of the full sky, or $\sim 5200$ deg$^2$. The mocks
include all galaxies down to $\log\lgal=8.66$ in the $r$-hand in solar
units\footnote{We use a SDSS $r$-band absolute magnitude of the sun of
  4.65.}, equivalent to $\mr=-17$. This is substantially more volume
than is achievable at this magnitude limit in SDSS, but matches the
depth of the ongoing Bright Galaxy Survey within the Dark Energy
Spectroscopic Instrument Survey (DESI-BGS; \citealt{desi_fdr}), as
well as the upcoming WAVES-Wide survey (\citealt{waves}).
 
The mocks test a variety of situations that could plausibly arise,
even though they are not necessarily representative of the actual SDSS
galaxy distribution. In \mocka, the galay-halo connection for red and
blue galaxies are substantially different, as is the resulting
measurements of $\lsat$---at fixed halo mass, the blue galaxies are
more massive than the red galaxies, while at fixed $\lgal$, $\lsat$
for red galaxies is higher. In \mockb, the galaxy-halo connections for
red and blue galaxies are different but the $\lsat$ measurements are
the same. In \mockum, the mock created from {\sc UniverseMachine}, the
galaxy-halo connections are the same, but the $\lsat$ measurements are
different.

The mock data on which our tests of the group finder are
based---namely, the projected correlation function and the
color-dependent $\lsat$ measurements---will be presented in \S
\ref{s.results} with the results of the tests.

\subsection{Total Satellite Luminosity} 
\label{s.mocks_lsat}

As discussed in the introduction, one of the main enhancements to the
standard group finder in the inclusion of information from $\lsat$
measurements. The number of subhalos within a halo, and thus the total
amount of light in satellite galaxies, scales nearly linearly with the
host halo mass. At low halo masses, where the average number of
spectroscopic galaxies within a group is expected to be $<2$, $\lsat$
brings extra constraining power on the halos of these galaxies.

The procedure for measuring $\lsat$ is detailed and tested in
\cite{tinker_etal:19_lsat}. Central galaxies from the spectroscopic MGS
sample are cross-correlated with galaxies within the deep imaging of
the DESI Legacy Imaging Survey (DLIS; \citealt{legacy_surveys}). The
DLIS data is $\sim 6$ magnitudes fainter in the $r$-band than the MGS
sample limit, making the imaging data capable of detecting satellite
galaxies around the faintest of MGS central galaxies. 

Without redshift information for the DLIS galaxies, satellites cannot
be found in individual halos. Rather, $\lsat$ is measured for stacked
samples of central galaxies by subtracting the estimated background
number of imaging galaxies projected along the line of sight. To
impose minimal priors on the measurement, $\lsat$ is evaluated in a
fixed comoving aperture around each central galaxy of 50 $\hkpc$. 

\cite{alpaslan_tinker:19} used these measurements to quantify the
correlation between $\lsat$ and secondary galaxy properties---i.e.,
size, velocity dispersion, color, morphology, etc---at fixed galaxy
stellar mass. If these parameters correlated only with $\mgal$ and not
with the host dark matter halo, then
no correlation would exist with $\lsat$, but correlations were found
with nearly all galaxy properties. An example of these results is
shown in Figure \ref{f.sigv_corr}, which we discuss further in \S
\ref{s.mocks_mockb}. One of the key results is that star-forming and
quiescent central galaxies had different $\lsat$ values at fixed
$\mgal$. These measurements were consistent with the lensing
measurements of \cite{mandelbaum_etal:16}, in which red galaxies
occupied higher mass halos than blue galaxies of the same $\mgal$. The
$\lsat$ results indicated that this bimodality, which is highly
significant in the weak lensing for galaxies with
$\mgal\gtrsim 10^{10.5}$ $\msol$, shrinks for less massive galaxies,
such that faint red and blue galaxies have the same $\lsat$ values.

These data will be critical in the self-calibration of the halo-based
group finder. We note that, in order to make these $\lsat$
measurements, a group catalog is first required in order to create a
sample of central galaxies. For the purposes of these mock tests, we
will use the true centrals for the $\lsat$ mock data. When
implementing on survey data, the process will have multiple
iterations: the first iteration will use the previous group finder,
and the second using the central galaxies from the first
implementation of the self-calibrated group finder.

\subsection{Mock A}
\label{s.mocks_mocka}

To assign luminosity to the halos and subhalos in \mocka, we use the
\cite{blanton_etal:05} luminosity function. The Blanton measurement is
corrected for incompleteness of low-luminosity galaxies caused by
their low surface brightness, which is critical for estimating $\lsat$
robustly. The completeness corrections increase the amplitude of the
luminosity function by a factor of 3 at r-band magnitudes of
$M_r-5\log h\sim -15$. We use a scatter in $\log\lgal$ at fixed
$\mpeak$, $\slogl$, of 0.2 dex.

Galaxies are assigned to be red and blue based on their luminosity
and whether they are central or satellite. We set the probability
that a given central galaxy is quenched to be

\begin{equation}
  \label{e.fqcen}
  P_{q}(\lgal|{\rm cen}) =
  0.5\left[1+{\rm erf}\left(\frac{\log\lgal-9.9}{0.7}\right)\right]
\end{equation}

\noindent where $q$ indicates ``quenched'' and erf is the error
function. This fitting function is calibrated to match the quenched
fraction of centrals from the \cite{tinker_etal:11} catalog. The
probability that a satellite galaxy is quenched is

\begin{equation}
  \label{e.fqsat}
  P_{q}(\lgal|{\rm sat}) =  0.4\left[1+{\rm erf}\left(\frac{\log\lgal-9.9}{0.7}\right)\right] + 0.2.
\end{equation}


\noindent This function, also comparable to current group catalog
results, sets a minimum quenched fraction of 20\% and ensures that the
quenched fraction of satellites is always higher than the quenched
fraction of centrals at all luminosities.

At this stage of construction, the LHRMs are the same for central
galaxies regardless of whether they are red or blue. As a simulacrum
of the \cite{mandelbaum_etal:16} and \cite{alpaslan_tinker:19}
observations, we increase the luminosities of the blue galaxies by a
luminosity-dependent factor,

\begin{equation}
  \label{e.lumshift}
 \Delta \log L_{\rm gal} = 0.5\left[1+{\rm erf}\left(\frac{\log L_{\rm
  gal}-10.0}{0.7}\right)\right].
\end{equation}

\noindent Eq.~(\ref{e.lumshift}) only makes bright blue galaxies
brighter, but leaves faint blue galaxies unchanged. From the point of
view of halo occupation, this makes blue galaxies brighter than red
galaxies in massive halos, while preserving the $\lsat$ observations
that blue rand red galaxies have the same luminosities in lower-mass
halos. Implementing Eq.~(\ref{e.lumshift}) on the mock alters the
overall luminosity function and quenched fractions to some degree, but
these modifications do not subvert the physical realism of the
mock---the quenched fraction still monotonically increases with
$\lgal$, and the quenched fraction of satellites is always higher than
that of centrals. The final galaxy-halo relations and $\fsat$ values
for this mock will be presented in \S \ref{s.results_mocka} and Figure
\ref{f.fsat_shmr_mocka}, when we show the group finder predictions for
these quantities.

Galaxy clustering in the mock is measured for red and blue subsamples
in bins of $\log\lgal$. To follow the convention of
\cite{zehavi_etal:11} and other SDSS analyses, we use bins of $r$-band
magnitude, with lower limits being $\mr=[-17, -18, -19, -20, -21]$,
with each bin being one magnitude wide. To estimate errors for the
$\wp$ measurements, we use the jackknife method, dividing the plane of
the sky into 25 roughly equal areas.

Even though the volume-limited mocks are constructed with $\mr<-17$,
the abundance matching calculation extends down to $\mr<-14$, which is
the magnitude limit of the $\lsat$ data. Each galaxy in the
mock-spectroscopic sample is also assigned a value of $\lsat$, using
the same aperture size as in the SDSS data. Because $\lsat$ requires
subhalos to be well-resolved at low masses, artificial disruption of
substructure is a concern. In tests with smaller-volume, higher
resolution simulations, we find a non-zero but small difference in
$\lsat$ for halos less massive than $\sim 10^{12}$ $\msol$
(\citealt{tinker_etal:19}), but in the group-finding process we do not
attempt to fit the absolute values of $\lsat$---rather, we fit the
{\it relative} values of $\lsat$ for red and blue galaxies at fixed
$\lgal$, $\lsatred/\lsatblue$. Using relative values will ameliorate
many possible systematics that may come into play; theoretical
systematics such as numerical resolution or artificial disruption of
satellites (e.g., \citealt{vandenbosch_etal:18b}) should cancel, and
observational systematics such as miscalibration between the SDSS and
DLIS photometry, or uncertainties in the background subtraction to
make the $\lsat$ measurements, will be attenuated as well.


\begin{figure*}
\plotone{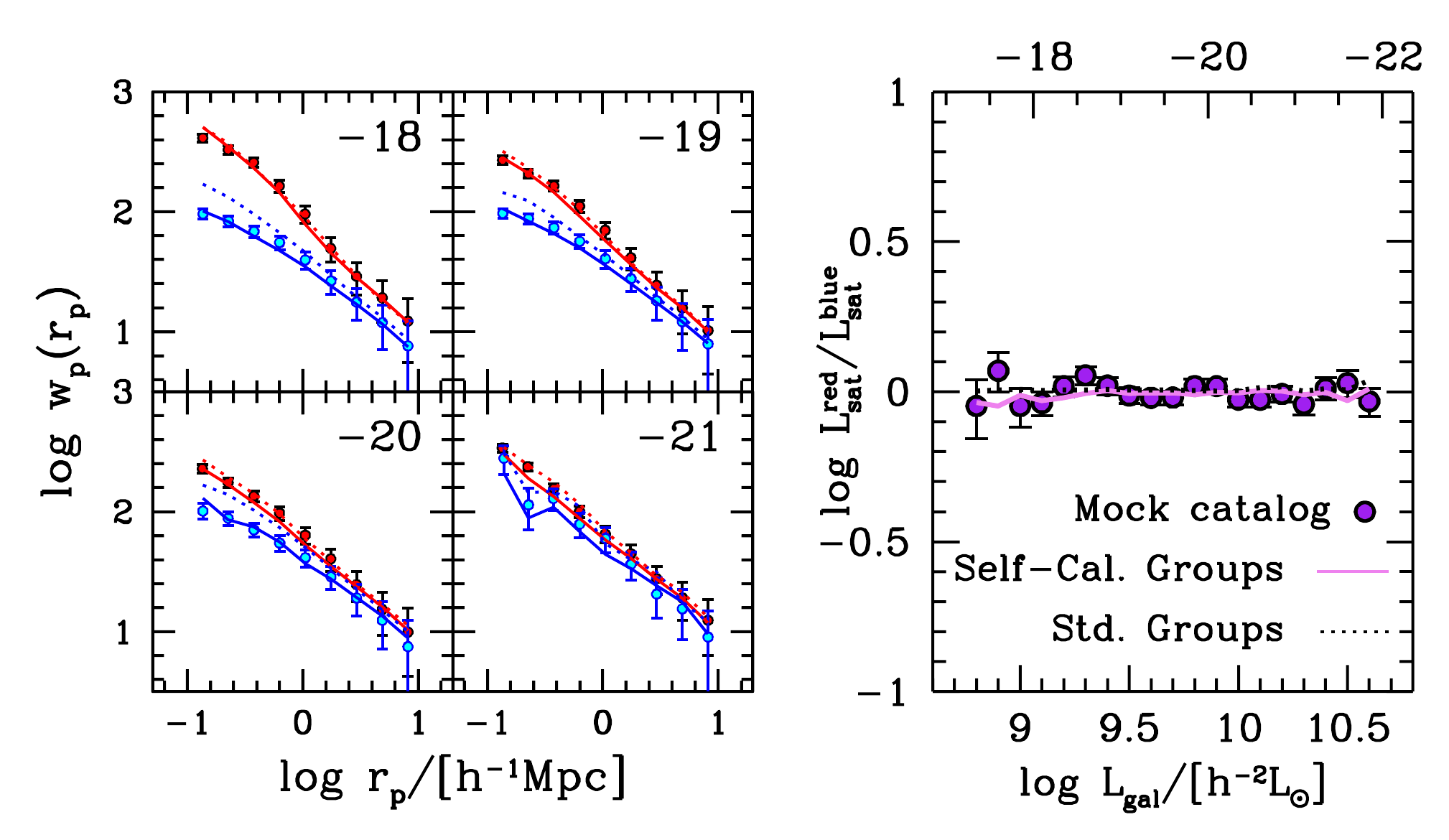}
\caption{ Observed properties of, and group finder fit to,
  \mockb. This figure is analogous to Figure
  \ref{f.wp_lsat_mocka}. The quartet of panels on left-hand side of
  the figure show the projected correlation function in bins of
  $r$-band magnitude, for red and blue galaxies. The measurements from
  the mocks are indicated by the points with error bars; red and blue
  colors indicate red and blue galaxies. The solid curves are the
  self-calibrated group catalog result. The dotted curves show the
  prediction from the standard group finder, without
  self-calibration. The large panel on the right-hand side shows the
  $\lsat$ ratio between red and blue galaxies at fixed $\lgal$. Points
  and curves are the same as the clustering panel. Because \mockb\ is
  constructed to have no difference in the observed $\lsat$ values
  between blue and red galaxies, the group finder fit and the
  prediction of standard group finder are nearly identical, and both
  are consistent with the mock data. \label{f.wp_lsat_mockb}}
\end{figure*}

\begin{figure*}
\plotone{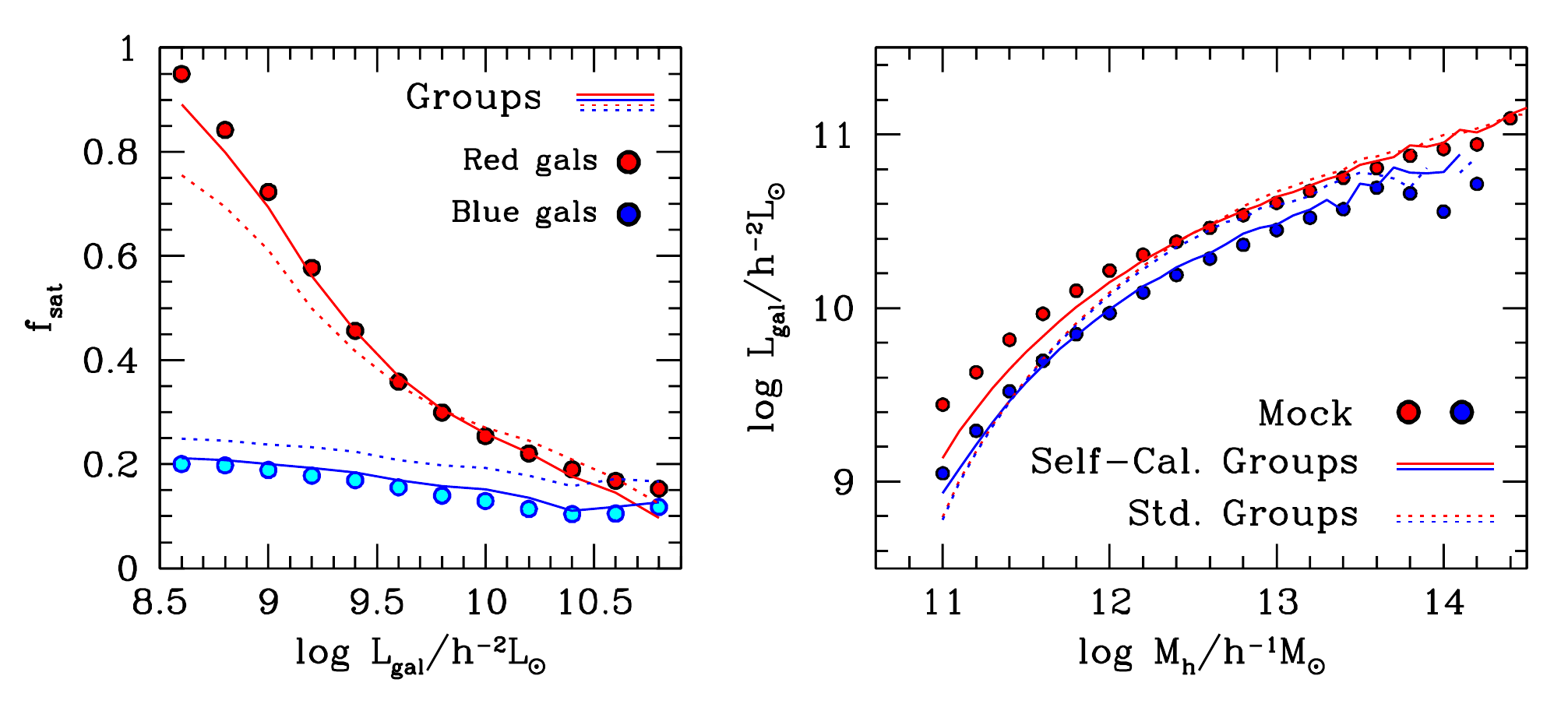}
\caption{ \label{f.fsat_shmr_mockb} Predicted quantities for \mockb.
  {\it Left-Hand Panel:} Satellite fraction, $\fsat$, as a function of
  $\lgal$, for red and blue galaxies. Colored circles show values from
  \mockb. Dotted curves show the predictions from the old group
  finder, which significantly underpredicts $\fsat$ for red
  galaxies. The solid curves show the predictions from the
  self-calibrated group finder. {\it Right-Hand Panel:} The LHMR for
  red and blue central galaxies. Filled circles show the values from
  \mockb. Solid curves show the predictions from the self-calibrated
  group finder. The inclusion of the $\chi$-$\lsat$ data, which we
  present in \ref{f.propx}, give the algorithm the constraining power
  to differentiate the LHMRs of red and blue galaxies. Dotted curves
  show the predictions of the standard group finder, which lie
  directly on top of each other. }
\end{figure*}

\subsection{MockB}
\label{s.mocks_mockb}

\mockb\ is constructed under the same framework as \mocka, but with
many substantive changes in order to test different aspects of the
group finder. First, we use the Sersic-profile $r$-band luminosity
function of \cite{bernardi_etal:13}, rather than the
\cite{blanton_etal:05} luminosity function. The Bernardi luminosity
function is consistent with Blanton at the knee of the luminosity
function, but yields a much higher abundance of luminous
galaxies. Thus the Bernardi LF yields a significantly different LHMR at the
massive end. The fit provided in Bernardi is not valid at lower
luminosities, thus we re-fit the Blanton two-Schechter functional form
to match the Blanton results at $\mr>-18$ and match the Bernardi
measurements and brighter magnitudes.

We incorporate a halo mass-dependent $\slogl$. In massive halos,
$\mhalo\gtrsim 10^{13}$ $\hmsol$, we set $\slogl=0.2$ dex, in
agreement with observations (\citealt{wechsler_tinker:18}). Below this
mass, $\slogl$ rises monotonically with decreasing $\log\mhalo$, to
$\slogl=0.35$ at $\mhalo=10^{11.0}$ $\hmsol$. This rise in scatter is
seen multiple hydrodynamic simulations and measured observationally by
\cite{cao_etal:19}. Using lensing results to put a lower bound on the
scatter at $\mhalo\approx 10^{12}$ $\hmsol$, \cite{taylor_etal:20}
also find that the scatter of $\mgal$ at fixed $\mhalo$ is
significantly higher than measurements at the group and cluster
scale. More than just incorporating these new observations, it is of
interest to determine if the self-calibrating group finder can recover
a mass-variable scatter.

The quenched fractions of central and satellite galaxies follow
Eqs. (\ref{e.fqcen}) and (\ref{e.fqsat}), without any shifting of
the luminosities for blue or red central galaxies. Thus, when binned
by stellar mass, the LHMRs of the two galaxy subsamples are the
same. However, the situation when binned by halo mass is quite different. At
fixed $\mhalo$, there is a distribution of $\lgal$. But the
Eq.~(\ref{e.fqcen}) depends on $\lgal$ only, thus quenched galaxies at
fixed $\mhalo$ are more likely to be brighter than average. Thus, when
plotted as a function of $\mhalo$, the LHMRs for red galaxies is
shifted above that for the blue sample.

As we will demonstrate, the standard group finder does not have the
constraining power to determine that the galaxy-halo connections are
different for red and blue galaxies in \mockb. Indeed, using clustering
information and the relative $\lsat$ values of of blue and red
galaxies does not provide sufficient constraining power. The last
observable included in the mocks, which we only incorporate in \mockb,
provides this information.

This observable is the correlation between $\lsat$ and a secondary
galaxy property at fixed $\lgal$ for central galaxies. Our motivation
for including these data come from the results of
\citealt{alpaslan_tinker:19}, which show strong correlations of $\lsat$
with secondary galaxy properties at fixed $\mgal$. An example of this
is shown in Figure \ref{f.sigv_corr}, for stellar velocity dispersion,
$\sigv$. The points with errors show the measurements for central
galaxies in SDSS. The steep slope implies that $\sigv$ holds
significant information about the halos around them. As a
demonstration of this, the solid curves show model curves that use
conditional abundance matching (\cite{hearin_watson:13,
  hearin_etal:14}) to match $\sigv$ to $\mhalo$ at fixed $\mgal$,
after abundance matching $\mgal$ onto $\mhalo$ with
$\slogm=0.2$. Different curves show how different correlations between
$\mhalo$ and $\sigv$ yield different slopes to the $\lsat$-$\sigv$
relation. To reproduce the observations, the coefficient of
determination, $r^2$, must be close to unity.

For the purposes of these tests, the galaxy property shall remain
hypothetical, referred to as $X$. Further, we make our mock
measurements as a function of the normalized parameter
$\chi\equiv (X-\bar{X})/\sigma_X$. As with the $\lsatred/\lsatblue$
measurements, we also only use $\lsat$ {\it relative} to the mean
value, i.e., $\lsatchibar$

As with the example in Figure \ref{f.sigv_corr}, we use conditional
abundance matching to incorporate a relationship between $\chi$ and
$\mhalo$. The method used here is taken from Appendix A in
\cite{tinker_etal:18_p3}. For convenience, we assume that $\chi$ is
normally distributed. In a narrow bin in $\log\lgal$, galaxies are
sorted in ascending order by $\mhalo$. We generate two Gaussian random
variables with zero mean and unit variance, $G_1$ and $G_2$. $G_1$ is
sorted in ascending order, while $G_2$ is left unsorted. For a given
galaxy in the sorted halo list, we assign $\chi$ by the weighted sum of
$G_1$ and $G_2$ such that

\begin{equation}
  \label{e.rsquare}
  \chi = G_1\times \frac{r^2}{\mathcal R} +
  G_2\times\frac{(1-r^4)^{1/2}}{\mathcal R}, 
\end{equation}
  
\noindent where $r$ is the correlation coefficient and ${\mathcal R}$
is a normalization constant to preserve the unit Gaussian distribution
of $\chi$. Because the scatter is lognormal in $\lgal$, it will be
approximately lognormal in $\mhalo$ at fixed $\lgal$, thus
Eq.~(\ref{e.rsquare}) correlates $\chi$ with $\log\mhalo$. If $r=1$,
then $\chi=G_1$ and the random variables is perfectly correlated with
$\log\mhalo$. If $r=0$, then $\chi=G_2$ and each galaxy is assigned a
$\chi$ randomly. For the mock, we measure $\lsatred/\lsatblue$ for
five bins in $\log\lgal$ 0.5 dex in width, with the bin centers at
$\log\lgal=[9.0, 9.5, 10.0, 10.5, 11.0]$. 

Our fiducial version of \mockb\ has $r^2=0.86$, which yields
correlations between $\chi$ and $\lsat$ that are consistent with that
seen in Figure \ref{f.sigv_corr}, and other properties investigated in
\cite{alpaslan_tinker:19}. We also construct versions of \mockb\ with
$r^2=1$ and $r^2=0.71$.

\subsection{Universe Machine Mock}

We use the public data release\footnote{\tt
  https://www.peterbehroozi.com/data.html} of {\sc UniverseMachine}
DR1 for the $z=0$ output of the Bolshoi-Planck simulation. These data
include halo mass, galaxy stellar mass, and star formation rate. We
therefore use $\mgal$ rather than $\lgal$ as the base galaxy property
to correlate with $\mhalo$. To separate galaxies in red and blue
subsamples, we use the specific star formation rate,
$\ssfr=\sfr/\mgal$, with a break point at $\log\ssfr=-11$, which
cleanly separates the two populations in SDSS data
(\citealt{wetzel_etal:12_groups1, hahn_etal:17_tq,
  tinker_etal:18_p3}).

The public data release doe not include luminosity information, thus
we assign values of $\lsat$ to each halo from the list of halos
produced for \mocka. For each UM halo, we find the 20 halos from
\mockb\ closest by their mass, and then select $\lsat$ randomly from
that list of 20. We check this measurement of $\lsatred/\lsatblue$ by
comparing it to summing the {\it stellar mass} in satellites within
the same projected separation as $\lsat$ measurements, and find that
$M_{\ast\rm sat}^{\rm red}/M_{\ast\rm sat}^{\rm blue}$ is consistent
with our approach. We will quantitatively compare the two in \S
\ref{s.results_um}.

From the UM DR1 data, we construct a volume-limited mock complete down
to $\mgal=10^{9.0}$ $\msol$. This yields a galaxy number density that
is 0.77 times that of the luminosity-defined mocks. Clustering of the
red and blue subsamples is measured in bins of $\log\mgal$ that are
0.5 dex wide, with the lower limit of each bin being
$\log\mgal=[9.0, 9.5, 10.0, 10.5, 11.0]$.

\mockum\ differs from the luminosity-defined mocks in several
important ways. First, although \mockum\ is built upon the
Bolshoi-Planck simulation, \cite{behroozi_etal:19} employ `orphan'
satellites to track galaxies that reside in subhalos that have become
numerically disrupted. Thus the overall number of satellite galaxies
will be higher in \mockum, as well as their spatial distribution
within their host halos. Second, \um\ uses empirical forward-modeling
to match observed galaxy statistics as a function of redshift,
tracking individual halos as they evolve across cosmic time. Thus,
galaxy quenching is not a simple function of $\lgal$ as in \mocka\ and
\mockb. Quenching probability correlates with halo properties,
including mass and formation history. 

As with \mocka, we do not incorporate the galaxy
secondary parameter, $\chi$.

\section{Group Finding Methods}
\label{s.methods}

Here we outline both the original halo-based algorithm, then describe
the extensions to this algorithm. We define halos as virialized,
spherical objects such that

\begin{equation}
\label{e.halodef}
M_\Delta = \frac{4}{3}\pi R^3_\Delta\Delta\bar{\rho}_m.
\end{equation}

\noindent where $\Delta$ is the mean overdensity of the halo and
$\bar{\rho}_m$ is the mean matter density of the universe in comoving
units. Thus, the comoving radius of a halo is independent of redshift
at fixed mass. In this definition, $\Delta$ is a free parameter. So
long as the choice is well-motivated and consistent throughout all
calculations, the results and insensitive to the specific value of
$\Delta$. Here we choose $\Delta=200$, thus $\mh=M_{200}$.

We further assume that all halos follow the universal density profile
of \cite{nfw:96} (NFW), which parameterizes the profile
through a single free parameter, referred to as the scale radius of
the halo, $r_s$;

\begin{equation}
\label{e.nfw}
\rho_h(r) = \frac{\rho_0}{(r/r_s)(1+r/r_s)^2}
\end{equation}

\noindent where $\rho_0$ is a normalization constant determined by the
mass of the halo. The concentration of a halo is the ratio of the halo
radius to the scale radius, $c_\Delta=R_\Delta/r_s$. Concentration and
mass are weakly correlated with each other, with a spread of
$\sim 0.12$ dex in $\log c$ at fixed $\mhalo$ (e.g.,
\citealt{bullock_etal:01}). We are not able to infer concentrations of
individual groups, thus we assume that all halos fall on the mean
concentration-mass relation of \cite{maccio_etal:08}.

From the virial theorem, the velocity dispersion within the dark
matter within a halo is

\begin{equation}
\label{e.veldisp}
\sigma_v^2 = \frac{GM_\Delta}{2R_\Delta}(1+z),
\end{equation}

\noindent where the factor of $(1+z)$ converts comoving radius
$R_\Delta$ to physical units. 

\subsection{Basic Algorithm for the Halo-Based Group Finder}
\label{s.basic}

The basic algorithm employed by \cite{tinker_etal:11} is presented
here, which follows \cite{yang_etal:05}, with minor modifications. We
expand on some of the steps in the subsequent text as necessary.

\begin{enumerate}
\item Inverse-abundance match to assign initial halo masses to
  galaxies, and label all galaxies as central.
  \vspace{-0.25cm}
\item Sort galaxies by halo mass in descending order.
  \vspace{-0.25cm}
\item In that order, for each central galaxy, determine the membership
  probability of neighboring (lower-luminosity) galaxies. If the
  probability is above a threshold value, assign that neighbor as a
  satellite to the central galaxy. If a central galaxy in the list has
  been re-classified as a satellite of a brighter galaxy, skip
  this step for that re-classified galaxy.
  \vspace{-0.25cm}
\item For each group, sum the total amount of luminosity (or stellar
  mass).
  \vspace{-0.25cm}
\item Sort the groups by total luminosity in descending order.
  \vspace{-0.25cm}
\item Assign halo masses to the groups by inverse-abundance matching.
  \vspace{-0.25cm}
\item If the value of $\fsat$ is converged, then exit. Otherwise, sort
  the central galaxies by halo mass and return to step 3.
  \vspace{-0.25cm}
\end{enumerate}

\noindent {\it 1.} Abundance matching is a technique to assign
galaxies to dark matter halos in simulations
(\citealt{kravtsov_etal:04, vale_ostriker:06, conroy_etal:06,
  wechsler_tinker:18}). Here, we proceed in the opposite
direction---we assign halo masses to galaxies. In this step, we use
the simplest form of abundance matching, expressed as

\begin{equation}
  \label{e.am}
  \int_{\lgal}^\infty \Phi(\lgal^\prime)\,d\lgal^\prime =
\int_{\mh}^\infty n(\mh^\prime)\,d\mh^{\prime},
\end{equation}

\noindent where $\Phi(\lgal)$ is the galaxy luminosity function and
$n(\mh)$ is the halo mass function. By matching the cumulative
abundances of galaxies and halos, Eq.~(\ref{e.am}) states that
galaxies of luminosity $\lgal$ reside in halos of mass $\mh$. This
implementation assumes no scatter between halo and galaxy properties,
which we do not have the power (yet) to constrain. When applying
Eq.~(\ref{e.am}) to all galaxies in a sample, the LHMR produced is a
poor approximation of the true relation at the the luminous end, but it
is accurate enough to serve as a starting point of the algorithm. It
also has the benefit of taking negligible computing time, in
comparison to friends-of-friends group finders or other percolation
methods. At this point in the algorithm, every galaxy is classified as
`central' for the purposes of group-finding.

\vspace{0.2cm}
\noindent {\it 3.} The probability that a neighbor galaxy is a member
of a given group is separated into two components; one based on the
projected separation between the neighbor and the central galaxy, and
the line-of-sight separation of the neighbor and central. The critical
insight of the \cite{yang_etal:05} algorithm is that we can bring in
our prior knowledge of dark matter halos from numerical
simulations to inform these choices.

We assume that satellite galaxies within a group are spatially
distributed the same as the dark matter within the halo. Thus, we can
use the NFW profile in Eq (\ref{e.nfw}) to quantify the relative
probability that a neighbor with separation $R_p$ is satellite within
a given halo. The projected probability, $P_{\rm proj}$, is equal to
the projected, normalized NFW density profile, expressed by
\cite{yang_etal:05} as

\begin{equation}
  \label{e.nfw_proj}
P_{\rm proj}(R_p)= 2r_s\bar{\delta}{f(R_p/r_s)}\,,
\end{equation}
where $R_p$ is the comoving projected separation between the central
galaxy and the candidate satellite galaxy, 
\begin{equation}
\label{fx}
f(x) = \left\{
\begin{array}{lll}
\frac{1}{x^{2}-1}\left(1-\frac{{\ln
{\frac{1+\sqrt{1-x^2}}{x}}}}{\sqrt{1-x^{2}}}\right)   &  \mbox{if   $x<1$}  \\
\frac{1}{3}   &   \mbox{if   $x=1$}   \\   
\frac{1}{x^{2}-1}\left(1-\frac{{\rm
      atan}\sqrt{x^2-1}}{\sqrt{x^{2}-1}}\right) & \mbox{if $x>1$}
\end{array} \right.\,,
\end{equation}
and 
\begin{equation}
\bar{\delta} = {200 \over 3} {c_{200}^3 \over {\rm ln}(1 + c_{200}) -
c_{200}/(1+c_{200})}. 
\end{equation}

For an isothermal, isotropic velocity distribution of satellites, with
velocity dispersion given by Eq.~(\ref{e.veldisp}), the line-of-sight
probability that a galaxy with redshift separation $\Delta z = z_{\rm
  cen}-z$ is

\begin{equation}
  \label{e.prad}
  P_z(\Delta z) =
  \frac{c}{\sqrt{2\pi}\sigma_v}\exp\left[\frac{-(c\Delta
        z)^2}{2\sigma_v^2}\right],
    \end{equation}

\noindent where $c$ is the speed of light. For a galaxy with projected 
separation $R_p$ and line-of-sight separation $\Delta z$ from a given
central galaxy, we set the probability that it is satellite as 
    
\begin{equation}
  \label{e.psat}
  P_{\rm sat} = \left[1-\left(1+P_{\rm proj}P_{z}/\bsat\right)^{-1}\right],
\end{equation}

\noindent where $\bsat$ is a free parameter. In the
\cite{yang_etal:05} implementation, this parameter was calibrated on
mocks, with a value of $\bsat=10$. This value was also used in the
\cite{tinker_etal:11} implementation. If $\psat>0.5$ then it is
classified as a satellite for that group. $\psat$ is not a true
probability, but rather a quantity that correlates with our confidence
the classification of a given galaxy. The utility of having a
continuous variable for $\psat$ is to construct `pure' samples of
central galaxies, where one has higher confidence that each galaxy in
the sample is indeed a true central, at the cost of some completeness
of the sample. 

\vspace{0.2cm}
\noindent {\it 6.} Inverse-abundance matching to assign halo mass to
total group luminosity requires a minor modification of
Eq.~(\ref{e.am}) to

\begin{equation}
  \label{e.amtot}
    \int_{\lgrp}^\infty n(\lgrp^\prime)\,d\lgrp^\prime =
\int_{\mh}^\infty n(\mh^\prime)\,d\mh^{\prime},
\end{equation}

\noindent where $n(\lgrp)$ is the abundance of the groups. 

\subsection{Extension and Self-Calibration}

To construct a self-calibrated group finder, we extend the halo-based
algorithm, and incorporate new data, in three key ways.

\vspace{0.2cm}
\noindent {\it 1. Allowing variability in $\bsat$:} The threshold for
assigning a galaxy to be a satellite within a group is set by
Eq.~(\ref{e.psat}). We first break $\bsat$ into separate values for
red and blue galaxies, $\bsatred$ and $\bsatblue$. Each of these
thresholds is then allowed to vary as a function of galaxy
luminosity:

\begin{equation}
  \label{e.bsat}
  \bsatc = \beta_{0,c} + \beta_{L,c}(\log\lgal-9.5),
\end{equation}

\noindent where the subscript $c={r,b}$ refers to the different color
samples, with $r$=red and $b$=blue.

\vspace{0.2cm}
\noindent {\it Constraining power from the data:} Eq.~(\ref{e.bsat})
gives us four free parameters, two for red galaxies and two for
blue. The data most useful for constraining these parameters are the
projected correlation functions of red and blue galaxies. $\wp$.  As
discussed in the introduction, if the group finder has correctly
assigned galaxies to halos, then the halo occupation produced by the
group finder should accurately predict the clustering of
galaxies. Two-point clustering, especially at scales of $r_p\lesssim
1$ $\hmpc$, is especially sensitive to the fraction of galaxies that
are satellites, $\fsat$. After fitting for $\wp$, to confirm 
that the group catalog is self-consistent, we check that the catalog
correctly reproduces the input value of $\fsat$ as a function of
$\lgal$ for mock red and blue galaxies.

\vspace{0.2cm}
\noindent {\it 2. Weighting the luminosity of red and blue central
  galaxies.} To account for the fact that red and blue central
galaxies may have different luminosities at fixed halo mass, we weight
the central galaxy luminosity when compiling the list of total group
luminosities used Eq.~(\ref{e.amtot}) to assign halo mass. The log of
the weight factor, $\wcen$ scales with luminosity as:

\begin{equation}
  \label{e.wcen}
  \log \wcenc =\frac{\omega_{0,c}}{2}\left[1+{\rm erf}\left(\frac{
      \log\lgal-\omega_{L,c}}{\sigma_{\omega, c}} \right)\right], 
\end{equation}

\noindent where $c=r$ or $b$ to indicate galaxy color. The form of
Eq.~(\ref{e.wcen}) affords maximum flexibility in the weight
factors. Positive and negative values of $\omega_0$ will either
upweight or downweight groups in the rank-ordered list. The
characteristic luminosity $\omega_L$ allows galaxies to be unweighted
at either low or high luminosities, and $\sigma_\omega$ allows the
weights to turn on abruptly, or to extend the transition so wide that
all galaxies in the sample receive equal up- or down-weight. We
implement Eq.~(\ref{e.wcen}) separately for red and blue central
galaxies, creating six free parameters. Having separate weights for
the two samples increases the flexibility further, allowing the
relative weights between the two to be non-monotonic. Because we are
less concerned with constraining the free parameters than constraining
the LHMRs, many degrees of freedom---which possibly create
degeneracies in the parameter space---is not a concern.

\vspace{0.2cm}
\noindent {\it Constraining power from the data:} Most of the
constraining power on $\wcenr$ and $\wcenb$ comes from measurements of
the $\lsatr/\lsatb$ ratio. This quantity is closely related to the
ratio of host halo masses at fixed $\lgal$. If this ratio deviates
from unity, then these galaxies occupy different halos. This does not
mean that the LHMRs must be different---it may be that the LHRMs are
the same, but different values of $\slogl$ may cause differences in
$\mhalo$ when binned by $\lgal$.

Although some inference on $\wcenr$ and $\wcenb$ comes from the
clustering data, the constraining power of $\wp$ for central galaxies
is limited. Measurements of clustering amplitude at large scales is
often used to infer halo mass, but for $\mhalo\lesssim 10^{13}$
$\hmsol$ in the local universe, the halo bias function flattens out,
thus it is difficult to infer much about central masses in the same
way that clustering can constrain satellite fractions.

\vspace{0.2cm}
\noindent {\it 3. Individual weights on based on secondary galaxy
  properties.} Figure \ref{f.sigv_corr}, demonstrates that individual
properties of galaxies contain information about their host halos at
fixed $\lgal$. Motivated by these results, we include weight factors
based on the hypothetical normalized galaxy property $\chi$, such that

\begin{equation}
  \label{e.wchi}
  w_\chi = \exp(\chi/\omega_\chi).
\end{equation}

\noindent We further separate these weight factors for red and blue
galaxies, $\wchir$ and $\wchib$, thus creating two new free
parameters, $\omega_{\chi,r}$ and $\omega_{\chi,b}$. The exponential
form of Eq.~(\ref{e.wchi}) is motivated by the log-linear results in
Figure \ref{f.sigv_corr}. We separate red
and blue galaxy samples because the correlations between galaxy
properties and $\lsat$ are usually different, and the properties
themselves can vary systematically between red and blue galaxies---for
example, in relation to Figure \ref{f.sigv_corr}, at fixed $\mgal$ blue
galaxies have lower $\sigma_v$ than red galaxies. (See the extensive
discussion in \citealt{alpaslan_tinker:19}).

\vspace{0,2cm}
\noindent {\it Constraining power from the data:} These two weight
factors are primarily constrained by the measurements described in
\mockb: $\lsatchibar$. Including the $\wchi$ weights will change the
scatter in the LHMR, and can thus alter predictions for the
$\lsatred/\lsatblue$ ratio, as well as influence the large-scale
galaxy bias for luminous samples of galaxies. But the influence of
these data is secondary.

\subsection{Group Finding and Fitting Procedure}

Our self-calibrated group finder has 12 free parameters. Four govern
satellite occupation, six control the relative LHRMs of red and blue
central galaxies, and the last two control individual weights on red
and blue central galaxies. These parameters are compiled together in
Table 1. To find the optimal parameters of the group
finder, we use the following data vector of observables from the
mocks described in \S \ref{s.mocks}:

\begin{itemize}
  \item $\wp$ for red and blue galaxies in 5 bins of magnitude. Each
    measurement has 9 bins in $r_p$, logarithmically spaced between
    $r_p=0.1$ and 10 $\hmpc$. 
  \item $\lsatred/\lsatblue$ for central galaxies, measured in 16 bins
    of $\log\lgal$. Error bars are estimated using the bootstrap
    resampling method.
  \item For \mockb\ only: $\lsatchibar$ for central galaxies,
    measured in 5 bins of $\log\lgal$. This is done separately for
    both red and blue galaxies.
    \item For the UM mock, the number of data points is the same as
      for \mocka, but $\wp$ and $\lsat$ quantities are binned in
      $\log\mgal$ as opposed to luminosity.
\end{itemize}

For a given set of parameters, the group finder is run using the
procedure outlined in the basic algorithm in \S \ref{s.basic}. The
only change in the basic algorithm is in the extra freedom in $\bsat$
and the total luminosity of the group that is used to rank-order the
groups in the abundance matching halo mass assignment. Under the new
algorithm, the group luminosity is

\begin{equation}
  \label{e.mgrp}
  L_{\rm grp} = L_{\rm cen}\times \wcenc\times \wchic + \sum_{i=1}^{N_{\rm
    sat}} L_{{\rm sat},i}
  \end{equation},

\noindent where $\wchi=1$ \mocka\ and \mockum. This yields a
label of `central' or `satellite' for each galaxy in the mock sample,
as well as an estimated halo mass. From this catalog, we make
predictions for all the quantities in the list above.
  
For $\wp$, we measure the mean halo occupation functions for red and
blue galaxies in the same magnitude bins for which $\wp$ is
measured. These HODs are tabulated separately for central and
satellite galaxies. The halos of our N-body simulation are then
populated according to these HODs, assuming a Poisson distribution for
satellites and a Bernoulli distribution for centrals.  For each color
and bin in $\mr$, we use the {\tt corrfunc} code to measure the group
finder's prediction for $\wp$ (\citealt{corrfunc}). We then calculate
the $\chi^2$ for each $\wp$ prediction.

For $\lsatred/\lsatblue$, we assign a value of $\lsat$ to each halo
from a tabulated list of the mean $\lsat$ as a function of $\mhalo$. This list
is constructed from the same N-body simulation used for $\wp$
prediction. We have tested assigning each group catalog halo a value
of $\lsat$  from an individual halo in the N-body simulation, but in
practice find that this only imparts noise in the prediction without
changing the mean. After each halo in the catalog has been
assigned $\lsat$, we separate the red and blue central galaxies and
measure $\lsatred/\lsatblue$ in the same manner as the mock
observations. 

For \mockb, we use the $\lsat$ values assigned to each halo in the
group catalog to measure $\lsat/\lsatbar$ in the same manner as in the
mock observations. The total $\chi^2$ for the data vector is then

\begin{equation}
  \label{e.chi2tot}
\chi^2_{\rm tot} = \sum_{\rm r,b}\sum_{i=1}^{N_{\rm
    bins}}\left[\chi^2_{w_p,i} + \chi^2_{\chi,i}\right] + \chi^2_{Lr/Lb}
\end{equation}

\noindent where $N_{\rm bins}=5$ is the number of luminosity bins,
noting once again that $\chi^2_{\chi}$ is only included for \mockb. We
use Powell's method (e.g., \citealt{press:92}) to minimize $\chitot$,
yielding the best-fit galaxy group catalog for each mock.

Although the primary results of this paper are based on the
Bolshoi-Planck simulation, we have found that the results of the group
finder are insensitive to which simulation is used to make the
predictions of the group catalog. We have performed tests using
different simulations to make the group-finder predictions, including the
original Bolshoi simulation, which has different initial
conditions and different cosmology, and the SMDPL which has a larger
volume as well as higher mass resolution.

\begin{figure}
  \epsscale{1.2}
  \hspace{-0.4cm}
\plotone{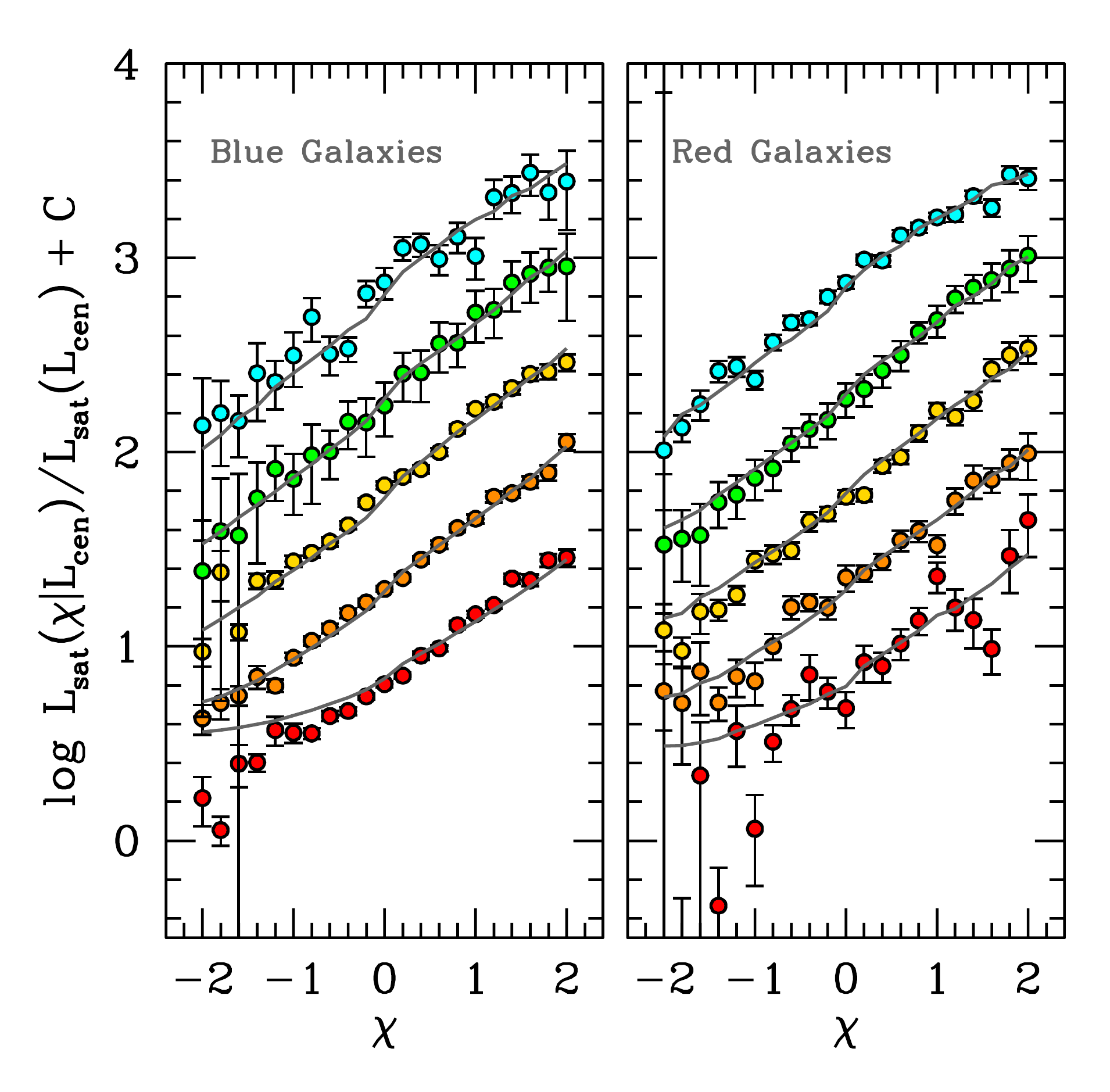}
\caption{ \label{f.propx} The correlation between $\lsat$ and a
  hypothetical normalized central galaxy property, $\chi$, in
  \mockb. This specific realization of \mockb\ has a coefficient of
  determination, $r^2$, of 0.87. Each set of points shows the mock
  measurements in bins of $\lgal$.  The $\lsat$ values are normalized
  by the mean value of $\lsat$ in the bin in $\lgal$. To differentiate
  the points within each panel, we have added an offset to each
  $\lgal$ bin. Error bars are estimated using jackknife
  resampling. In each panel, the solid curves show the results of
  the best-fit self-calibrated group finder. }
\end{figure}

\begin{figure}
  \epsscale{1.2}
  \hspace{-0.4cm}
\plotone{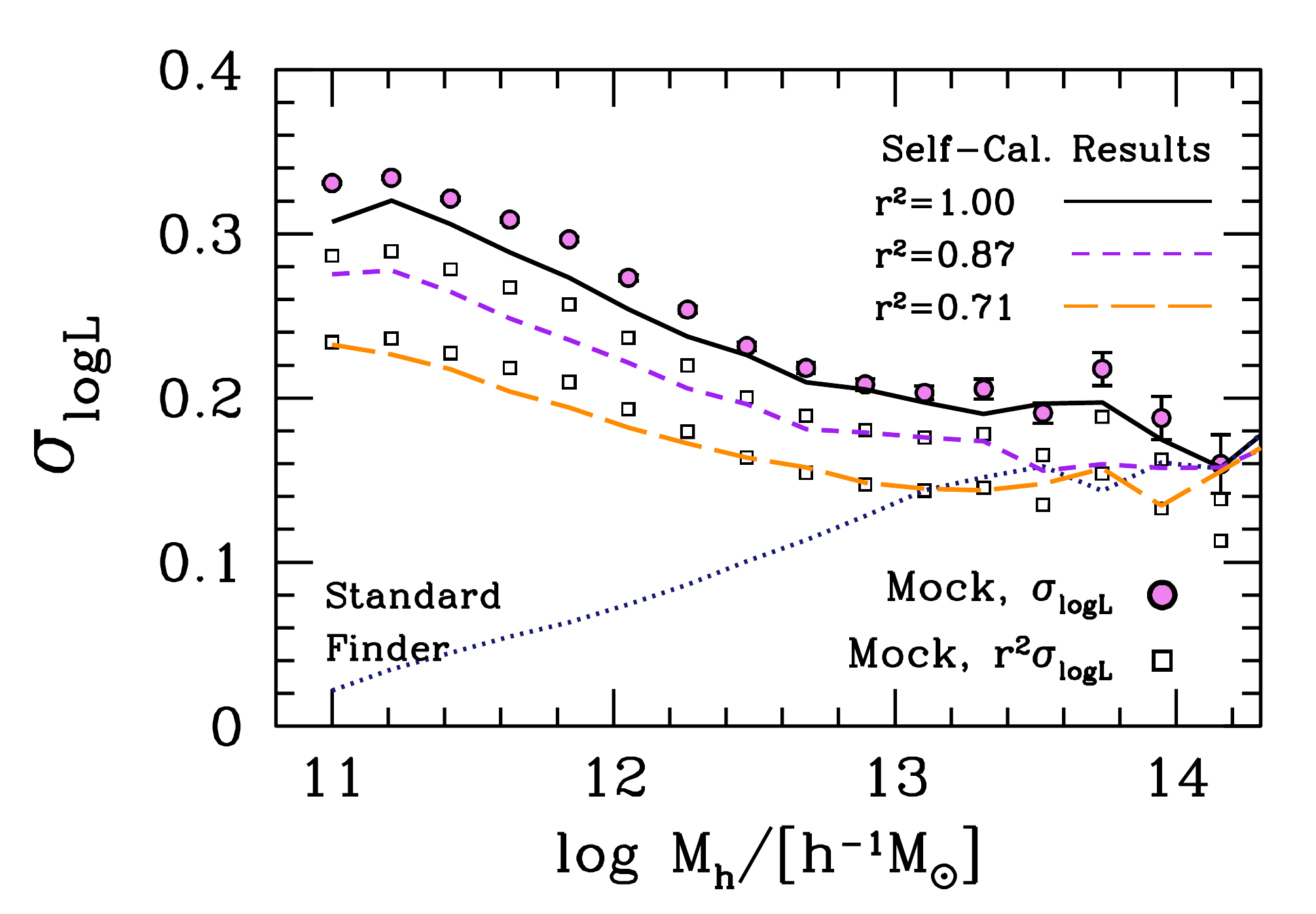}
\caption{ \label{f.scatter} The scatter of $\log\lgal$ at fixed
  $\mhalo$, $\slogl$. The filled circles show $\slogl$ in \mockb. The dotted
  curve shows the results of the standard group finder, which is not
  able to differentiate halos of central galaxies with the same
  luminosity when only the central is the only galaxy in the
  group. The other curves show the results of the self-calibrated
  group finder applied to different versions of \mockb\ where the
  value of $r^2$ is varied. The open squares show the scatter in
  \mockb\ multiplied by $r^2$.}
\end{figure}

\begin{figure}
  \epsscale{1.2}
  \hspace{-0.4cm}
\plotone{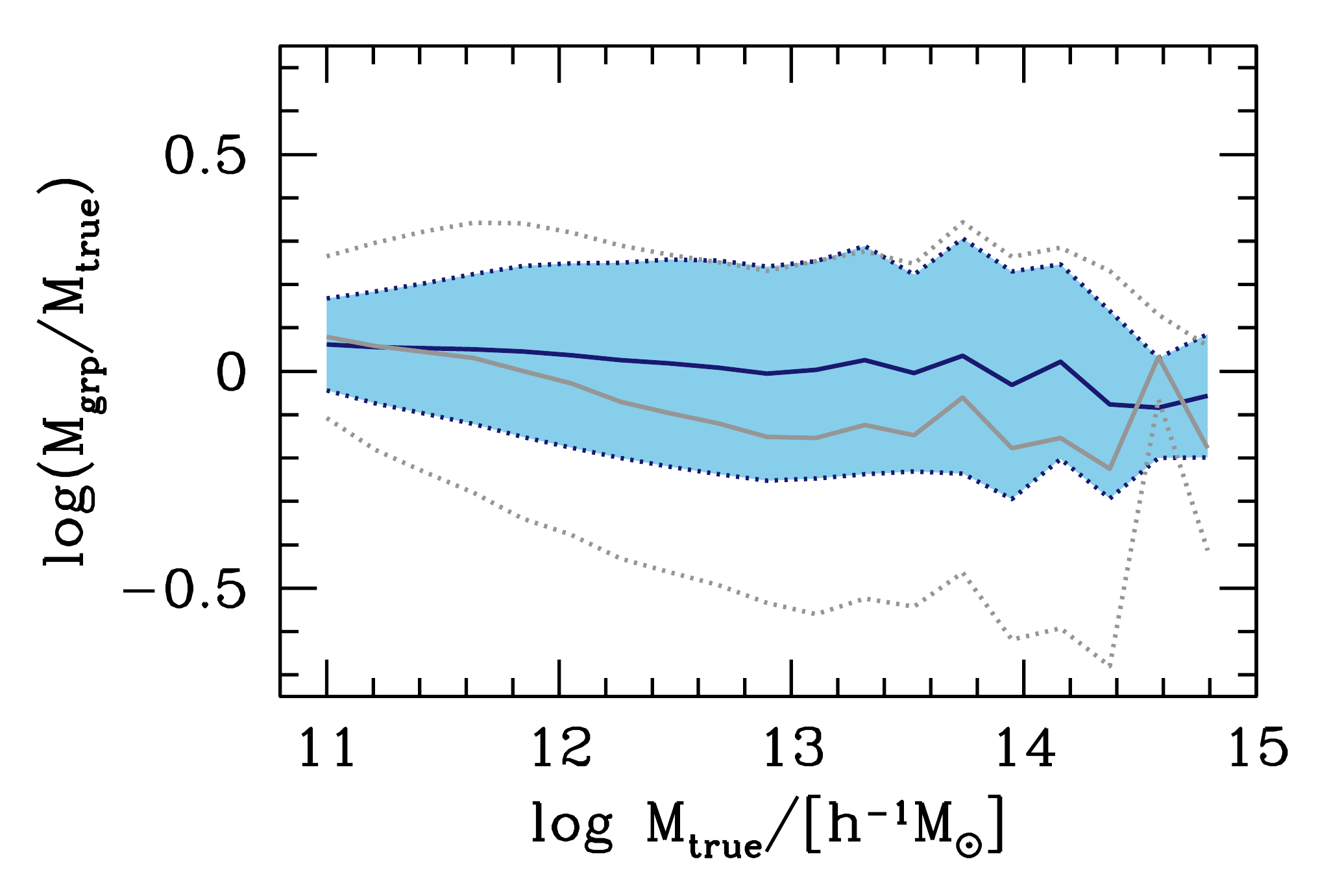}
\caption{ \label{f.mass_err} Errors in halo mass from the analysis of
  \mockb\ with $r^2=0.87$. The blue cuves show the results from the
  self-calibrated group finder, where the solid curve is the mean halo
  mass and the dotted curves indicate the inner-68\% range of
  values. The results from the standard group finder are shown with
  the gray curves, indicating that the self-calibrated algorithm
  yields roughly a factor of two improvement in the accuracy of
  $\log \mhalo$. The inclusion of the $\chi$-$\lsat$ data from Figure
  \ref{f.propx} provides the information that improves these
  constraints. Increasing $r^2$ shrinks the shaded region. Increasing
  $\slogl$ in the mock will increase the errors for both algorithms.}
\end{figure}

 \begin{deluxetable*}{crlccc}
\tablenum{1}
\tablecaption{Parameters of the Self-Calibrated Group Finder\label{tab:messier}}
\tablewidth{0pt}
\tablehead{
\colhead{Parameter} & \colhead{Purpose} & Equation & \colhead{\mocka} & \colhead{\mockb} &
\colhead{\mockum}  }
\startdata
$\beta_{0,b}$ & blue satellite threshold & Eq.~(\ref{e.bsat})---$\bsatblue$ & 16.56
& 15.07 & 15.81 \\
$\beta_{L,b}$ & blue satellite threshold & Eq.~(\ref{e.bsat})---$\bsatblue$ & -0.01
& 0.31 & 1.36 \\
$\beta_{0,b}$ & red satellite threshold & Eq.~(\ref{e.bsat})---$\bsatred$ & 7.74
& 9.46 & 9.86 \\
$\beta_{L,b}$ & red satellite threshold & Eq.~(\ref{e.bsat})---$\bsatred$ & 5.58
& 5.14 & 6.36 \\
$\omega_{0,b}$  & weight blue centrals & Eq.~(\ref{e.wcen})---$\wcenb$ & -1.13 & 0.76 & -1.11 \\
$\omega_{L,b}$  & weight blue centrals & Eq.~(\ref{e.wcen})---$\wcenb$ & 10.39  &
10.94 & 10.62 \\
$\sigma_{\omega,b}$  & weight blue centrals & Eq.~(\ref{e.wcen})---$\wcenb$ &
0.93 & 0.78 & 0.73 \\
$\omega_{0,r}$  & weight red centrals & Eq.~(\ref{e.wcen})---$\wcenr$ & -0.93 & 0.66 & -0.78 \\
$\omega_{L,r}$  & weight red centrals & Eq.~(\ref{e.wcen})---$\wcenr$ & 10.44 & 10.92 & 10.46 \\
$\sigma_{\omega,r}$  & weight red centrals & Eq.~(\ref{e.wcen})---$\wcenr$ & 0.55 & 0.84 & 0.56 \\
$\omega_{\chi,b}$ & $\chi$-property weight & Eq.~(\ref{e.wchi})---$\wchib$ & --- & 1.60 & --- \\
$\omega_{\chi,r}$ & $\chi$-property weight & Eq.~(\ref{e.wchi})---$\wchir$ & --- & 1.52 & --- \\
\enddata

\tablecomments{\mockb\ is the only mock that includes the measurements
  of the correlation between $\lsat$ and the normalized galaxy
  property, $\chi$. Thus the values of $\omega_{\chi,r}$ and
  $\omega_{\chi,b}$ are only listed for the fiducial version of this
  mock, which has $r^2=0.87$. }
\end{deluxetable*}

\section{Mock Results and Results on Mocks}
\label{s.results}

Before discussing the performance of the self-calibrated group finder,
we first describe the general characteristics of \mocka\ and the
performance of the previous halo based group finder, outlined in \S
\ref{s.basic}. We refer to the previous algorithm as the
`standard group finder.' We focus on the observable quantities---$\wp$
and $\lsat$---and the predicted quantities of $\fsat$ and the LHMR.

The left-hand side of Figure \ref{f.wp_lsat_mocka} is a quartet of
panels plotting the color-dependent clustering in bins of
magnitude. Consistent with SDSS results (\citealt{zehavi_etal:11}),
the red galaxies in the mock are significantly more clustered than the
blue galaxies. This is due to the higher $\fsat$ for red galaxies
relative to blue galaxies. On the right-hand side of the Figure is the
ratio of the $\lsat$ values for red and blue central galaxies. As
discussed in \S \ref{s.mocks}, the blue galaxies in \mocka\ are
brighter than the red galaxies at fixed halo mass. Thus, when binning
by the $\lgal$, the halo masses---and therefore the $\lsat$
values---for the red galaxies are higher.

Figure \ref{f.wp_lsat_mocka} highlights several of the issues of the
standard group finder discussed in the introduction. The dotted curves
in all panels show the results of applying the standard group finder
to \mocka, with $\bsat=10$.  For blue galaxies, the halo occupation
inferred by the standard finder yields clustering that is in good
agreement with the clustering of the mock catalog, but the halo
occupation of the red galaxies underpredicts the clustering by a
significant amount. In the right-hand panel, the standard group finder
yields a $\lsat^{\rm red}/\lsat^{\rm blue}$ ratio that is nearly unity
across the full range of central galaxy luminosity, even though the
blue galaxies in the mock reside in more massive galaxies. In the
mock, the red central galaxies will have more spectroscopic satellite
galaxies than the blue central galaxies at fixed $\lgal$, but this
effect by itself is not enough to yield the proper halo masses.

The color-dependent satellite fractions of the mock, $\fsat$, are
shown in the left-hand side of Figure \ref{f.fsat_shmr_mocka}. For
blue galaxies, $\fsat$ is largely constant with $\lgal$ until galaxies
get brighter than the knee in the luminosity function
$(\lgal\sim 10^{10.04}$ $\lsolhh$, according to the
\citealt{blanton_etal:03} results). $\fsat$ for red galaxies
monotonically increases with decreasing $\lgal$, reaching $\sim 90\%$
at $\lgal\approx 10^{8.7}$ $\lsolhh$. These results
are consistent with the HOD fitting of SDSS clustering in
\cite{zehavi_etal:11}. Here we see the reason that the standard group
finder woefully underpredicts the small-scale clustering of red
galaxies: the predicted $\fsat$ is much lower than the mock value,
never getting higher than $\sim 60\%$. The $\fsat$ for blue galaxies
is slightly higher than the mock value---the overall $\fsat$ (not
plotted) is in good agreement with the standard mock, but this
agreement is due to a cancellation of errors by overpredicting the
number of blue satellites and underpredicting the number of red
satellites. At low luminosities most galaxies are blue, so the
discrepancy is much larger for the red sample.

The right-hand side of Figure \ref{f.fsat_shmr_mocka} shows the
LHMRs for red and blue galaxies. The mean $\lgal$ for
blue and red central galaxies diverge at $\mhalo\gtrsim 10^{11.5}$
$\hmsol$, but because the central galaxy dominates the total
luminosity in the group, the standard finder has no way of determining
this. Thus, the standard finder's LHMRs for red and blue central
galaxies only separate at much higher $\mhalo$, once the number of
satellite galaxies in the halos dominates the total luminosity in the
halo. These problems are ameliorated substantially in the
self-calibrated algorithm, the results of which we show presently.

\subsection{MockA}
\label{s.results_mocka}

The solid curves in both sides of Figure \ref{f.wp_lsat_mocka} show a
markedly improved comparison to the mock data relative to the results
from the standard group finder. In the left-hand side, showing the
color-dependent clustering, the improved fit by the self-calibrated
group finder is due mainly to the extra freedom in the thresholds that
determine inclusion as a satellite in a larger group. The best-fit
parameters, listed in Table 1, indicate that $\bsatred$ is much lower
than the fiducial value of 10, with a significant positive slope with
$\log\lgal$. $\bsatblue$ is still constant with luminosity, but the
best-fit value is 16.6. Both of these adjustments yield a good fit to
the clustering, as well as an improved comparison to the satellite
fractions in the left-hand side of Figure \ref{f.fsat_shmr_mocka}.

The self-calibrated finder also yields a good fit to the $\lsat$ data
in the right-hand panel of Figure \ref{f.wp_lsat_mocka}. This is due
to the weights, $\wcenr$, and $\wcenb$, that are applied to the
luminosities of the central galaxies when rank-ordering the groups for
halo mass assignment. Figure \ref{f.bestfit} shows the ratio of
$\wcenb/\wcenr$ for the best-fit model. The difference at the low-mass
end, $\mhalo\lesssim 10^{11.5}$, is due mainly to how the samples are
binned; the mock sample has a hard cutoff at $\lgal=10^{8.6}$
$\lsolhh$, while the group catalog yields a cutoff at fixed halo
mass. Thus, even though the intrinsic relations may be equivalent in
the mock and the catalog, the binned quantities can diverge when the
bins encompass the limits of the sample.

\begin{figure}
  \epsscale{1.2}
\plotone{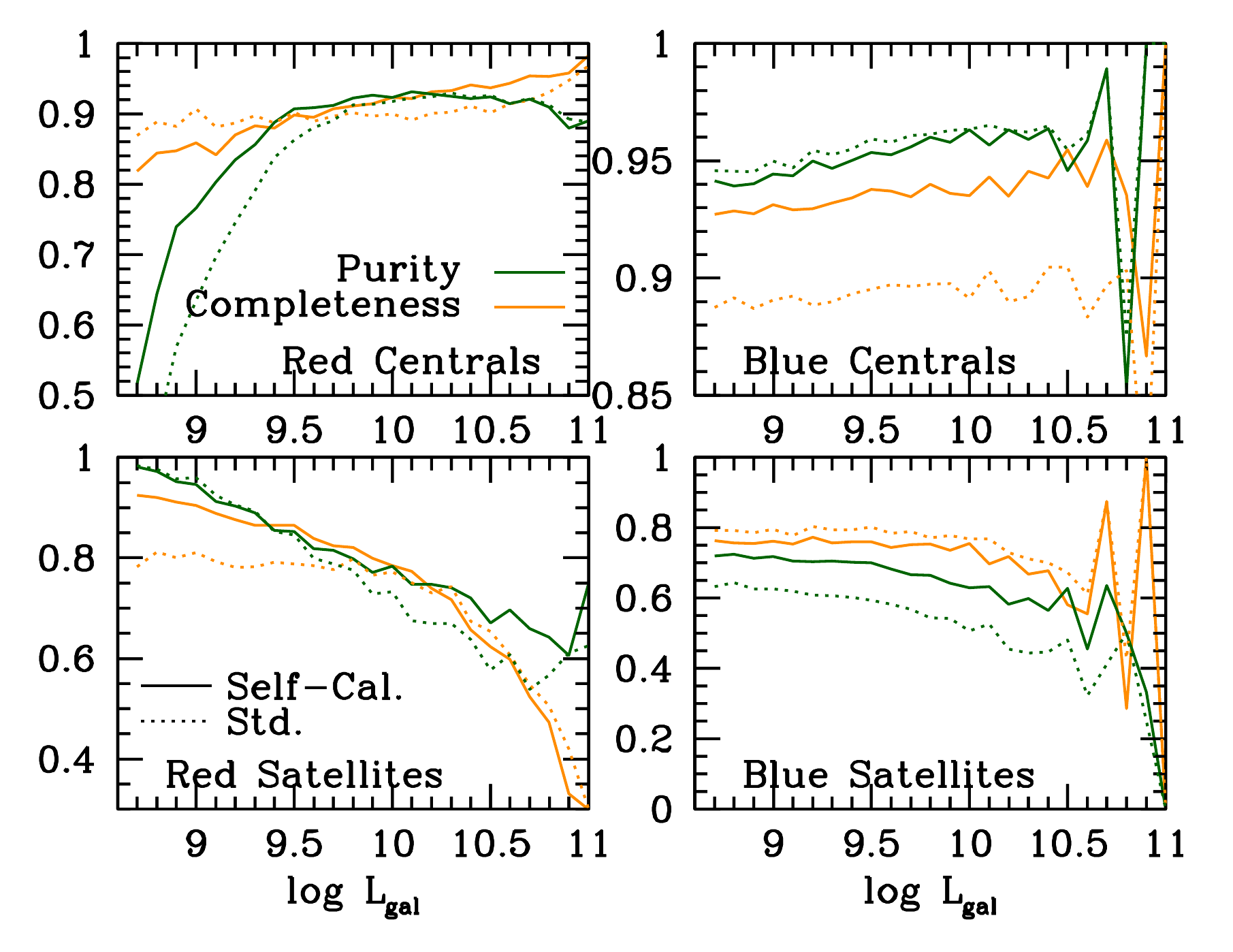}
\caption { \label{f.comp} Purity and completeness for group finders
  applied to \mockb. Solid curves indicate results results from the
  self-calibrated algorithm while dotted curves show the results from
  the standard group finder. Results are broken down into four galaxy
  classifications: red centrals, blue centrals, red satellites, and
  blue satellites. Completeness is defined as the fraction of galaxies
  in a given classification in the mock that have the same
  classification in the group catalog. Purity is the fraction of
  galaxies in a given classification in the group catalog that have
  the same classification in the mock. Note the change in $y$-scale in
  each of the panels. 
}
\end{figure}

\begin{figure}
  \epsscale{1.2}
\plotone{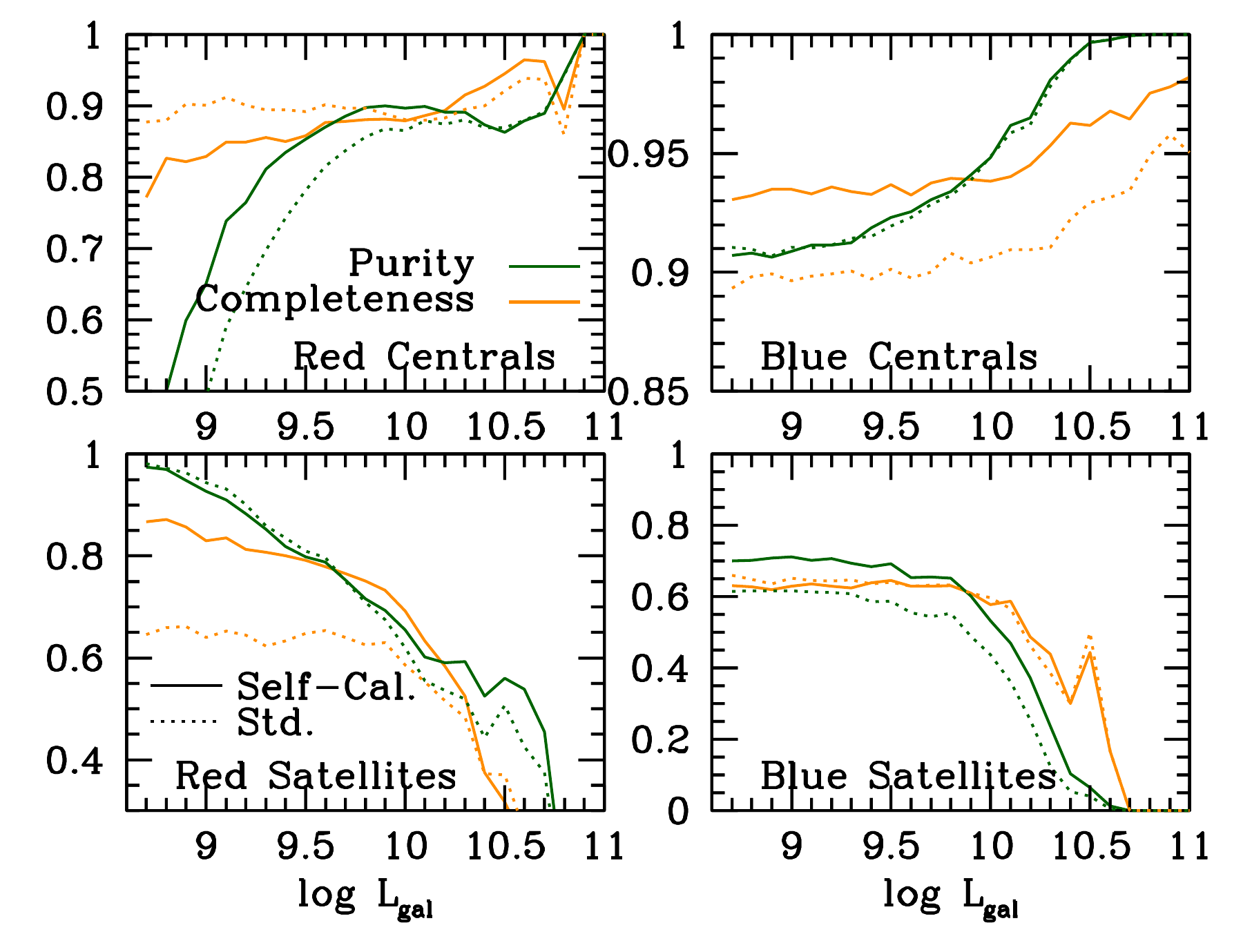}
\caption { \label{f.comp2} Purity and completeness for group finders
  applied to \mocka. All curves and definitions are the same as Figure
  \ref{f.comp}.
}
\end{figure}


\begin{figure*}]
\plotone{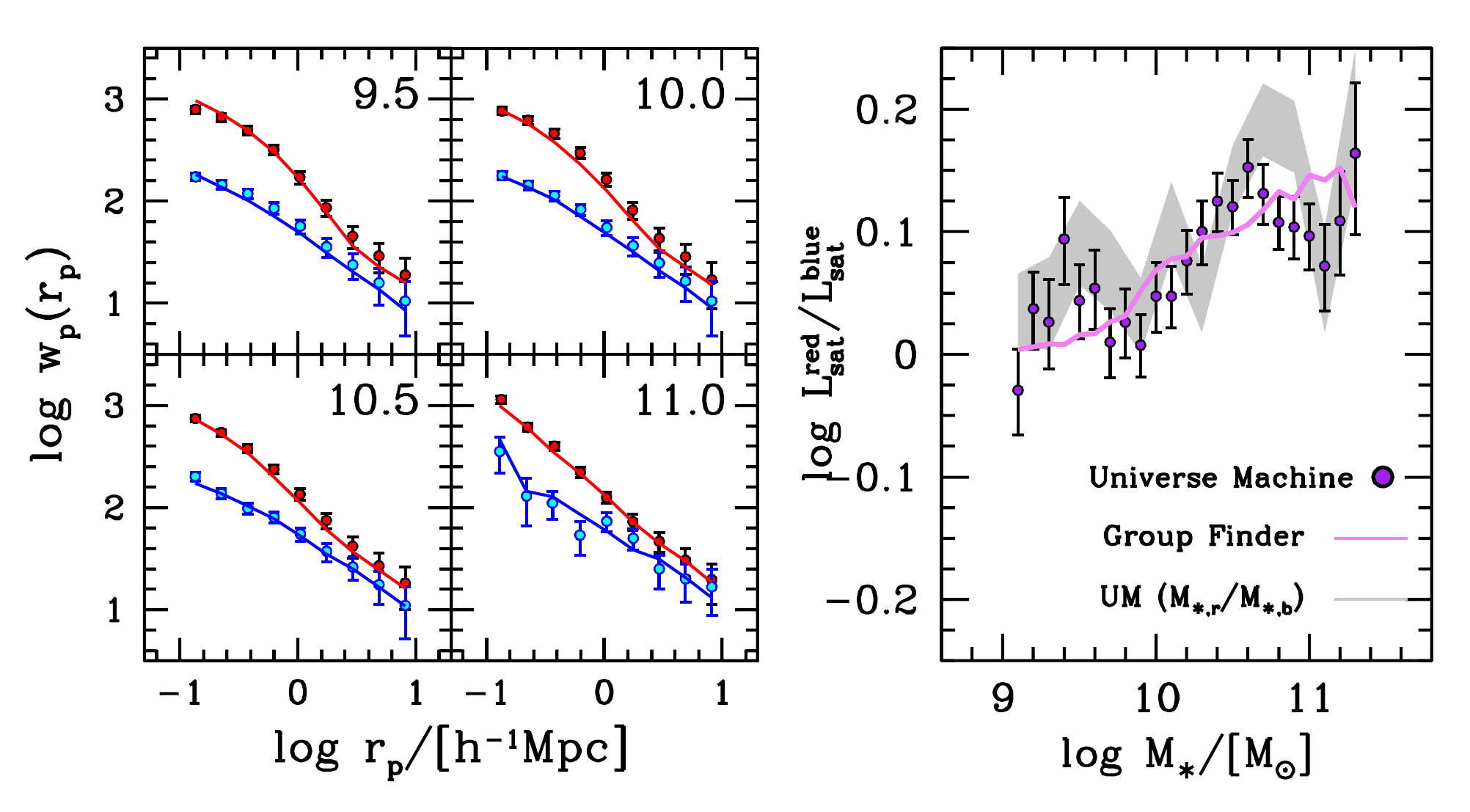}
\caption{ Observed properties of, and group finder fit to,
  \mockum. The quartet of panels on left-hand side of the figure show
  the projected correlation function in bins of $\log\mgal$,
  for red and blue galaxies. The measurements from the mocks are
  indicated by the points with error bars; red and blue colors
  indicate red and blue galaxies. The solid curves are the best-fit
  group catalog result. The large panel on the
  right-hand side shows the $\lsat$ ratio between red and blue
  galaxies at fixed $\lgal$. Points and curves are the same as the
  clustering panel.  \label{f.wp_lsat_um}}
\end{figure*}

\subsection{\mockb : $\fsat$ and LHMR}
\label{s.results_mockb}

The clustering and $\lsat$ observables for \mockb\ are shown in Figure
\ref{f.wp_lsat_mockb}. While the clustering is very similar to that of
\mocka, the $\lsat$ values are quite different, with the $\lsat$ ratio
being equal to unity across the full range of central galaxy
luminosity. The satellite fractions and LHMRs for red and blue
galaxies in the mock are shown in Figure \ref{f.fsat_shmr_mockb}.  The
satellite fractions are similar to \mocka, but the LHMR for red
galaxies is slightly higher than that for blue galaxies at all halo
masses. The reason for the difference is discussed in \S
\ref{s.mocks_mockb}, due to the criteria by which galaxies are
assigned their colors.

The results of the standard group finder are shown in each panel in
both figures with the dotted lines. In contrast to \mocka, the
clustering of red galaxies is well-fit by the standard
finder. However, now the clustering of blue galaxies is significantly
higher than the mock observables. The source of the discrepancy with
the clustering of blue galaxies is found in the left-hand side of
Figure \ref{f.fsat_shmr_mockb}; the satellite fraction of blue
galaxies is $\sim 25\%$ higher than the mock value of $0.20$. The
dramatic difference in the clustering highlights the sensitivity of
$\wp$ to $\fsat$.

Because the mock is constructed to have $\lsat$ values the same for
red and blue central galaxies, the standard finder yields a good fit
to the $\lsatred/\lsatblue$ observable, shown in Figure
\ref{f.wp_lsat_mockb}. In the right-hand side of Figure
\ref{f.fsat_shmr_mockb}, the standard finder yields LHMRs for red and
blue galaxies that are virtually identical. The shape of both LHMRs is
somewhat steeper than the mock values as well. This is due to the
increasing fraction of red galaxies with $\lgal$; at low $\lgal$, most
galaxies are blue and thus both LHMRs track that of the blue LHMR of
the mock. At higher $\lgal$, most galaxies are red, thus the dotted
lines track the red LHMR.

In both figures, the result of the self-calibrated galaxy group finder
are shown with the solid lines. \mockb\ has the addition of the
$\chi$-parameter, and its correlation with $\lsat$. Figure
\ref{f.propx} shows the correlation between $\chi$ and $\lsat$, in
bins of $\lgal$, broken into red and blue galaxies. In each bin of
$\lgal$, the results are normalized by the mean $\lsat$ within the
bin. For plotting purposes, we offset each $\lgal$ bin by a constant
that increases monotonically with bin luminosity. The curves show the
best-fit model from the self-calibrated group finder. Recall that the
coefficient of determination is $r^2=0.87$ for this model, which
yields correlations of $\lsat$ with $\chi$ that are consistent with
the \cite{alpaslan_tinker:19} results. The best-fit weight parameters
are $\omegachired=1.5$ and $\omegachiblue=1.6$. These values vary
inversely  with $r^2$, as larger values of $\omega_\chi$ reduce the
$w_\chi$ weights in Eq.~(\ref{e.wchi}). For the versions of \mockb\
with $r^2=1.0$ and $0.71$, the values of $\omega_\chi$ are 1.3 and
1.9, respectively, for both red and blue galaxies.

Figure \ref{f.fsat_shmr_mockb} compares the self-calibrated group
finder's prediction for the LHMRs to the actual values in the
mock. The prediction correctly separates the LHMRs between the two,
with the red galaxies being $\sim 0.2$ dex higher in luminosity at
fixed $\mhalo$.  It is the inclusion of the $\chi$-$\lsat$ data that
allows the self-calibrated group finder to correctly determine that
the LHMRs for red and blue galaxies are distinct. The
$\lsatred/\lsatblue$ data are equal to unity, and are well-fit by the
standard group finder. But the $\chi$-$\lsat$ data provides
information on the scatter of $\lgal$ at fixed $\mhalo$, which drives
the difference in the LHMRs. We discuss scatter in more detail
presently.

\subsection{\mockb: Scatter and Mass Errors}
\label{s.results_scatter}

The addition of the $\chi$ parameter in the model fitting allows us to
differentiate the halo masses of galaxies in halos that have no
spectroscopic satellites. Thus, this additional freedom makes the new
group finder able to estimate the scatter of $\log\lgal$ at fixed
$\mhalo$, $\slogl$. Figure \ref{f.scatter} shows the true scatter in
\mockb, as well as the results of various implementations of the
self-calibrating group finder. As described in \S \ref{s.mocks}, the
input scatter is variable, being consistent with observational
constraints of $\slogl\sim 0.2$ at high masses, and increasing to
$\sim 0.35$ at low $\mhalo$. The scatter estimated by the standard
group finder is shown with the dotted black curve. At $\mhalo\gtrsim
10^{13}$ $\hmsol$, it estimates a value of $\slogl$ $\sim 30\%$ below
the true value, consistent with results from
\cite{reddick_etal:13} and \cite{cao_etal:19}. However, as also shown in
these works, the scatter in the standard group catalog approaches zero
as $\mh$ approaches the minimum halo mass in the catalog.

The results of the self-calibrating group finder are shown with the
other curves in Figure \ref{f.scatter}. Here, we show the results for
three different versions of \mockb, with values of $r^2=0.71$, 0.875 and
1.0. The true $\slogl$---and all other properties of the mock---do not
change; only the value of $r^2$ and thus the slope of the correlation
between $\chi$ and $\lsat$ (e.g., Figure \ref{f.sigv_corr}. When
$r^2=1.0$, i.e., the normalized galaxy property $\chi$ correlates
perfectly with the halo mass, the new finder recovers the true scatter
of the mock. When $r^2$ is below unity, the variance in $\chi$ no
longer encodes all the information about the variance in $\mhalo$,
thus the new finder's estimate of $\slogl$ is biased low relative to
the true value. Figure \ref{f.scatter} shows that that scatter
inferred by the group finder is equivalent to $r^2\times \slogl$.

The inclusion of the $\chi$-$\lsat$ data increases the finder's
ability to accurately estimate halo masses themselves. Figure
\ref{f.mass_err} shows the ratio of the halo mass assigned by the
group finder to the true halo mass within the mock,
$\log(M_{\rm grp}/M_{\rm true})$, as a function of $M_{\rm true}$. At
$10^{13}$ $\hmsol$, the standard deviation is $\sim 0.2$ dex, which is
roughly a factor of two improvement over the standard group
finder. The self-calibrated algorithm also yields unbiased halo
masses, correcting the small bias of $\sim 0.1-0.15$ dex in the
standard finder.

\subsection{Purity and Completeness}

Figure \ref{f.comp} compares the purity and completeness in the
self-calibrated group catalog to the standard group catalog when
applied to \mockb. We calculate these quantities for four
classifications of galaxies: red central galaxies, blue central
galaxies, red satellite galaxies, and blue satellite galaxies. We
define completeness as the fraction of galaxies in original mock that
have the same classification in the group catalog. We define purity as
the fraction of galaxies in the group catalog that have the same
classification in the original mock.

Although the self-calibrated group finder shows higher purity and
completeness relative to the standard finder, we not that this is not
always the case. The standard finder outperforms by $\sim 5\%$ the
self-calibrated algorithm in the completeness of low-luminosity red
central galaxies. This is a result of the enhanced purity of red
centrals in the self-calibrated catalog, which yields an improvement
of $\sim 15\%$ in that quantity. There is a competition between purity
and completeness, especially when the subsample in question---red
centrals---is the smallest category.

The self-calibrated group finder yields most notable improvements in
the completeness of blue centrals, the completeness of red satellites
and low luminosities and the purity within the same classification at
high luminosities, and the purity of blue satellites at all
masses. The competition between purity and completeness causes the
self-calibrated algorithm to do slightly worse than the standard
approach in the purity of blue centrals and the completeness of blue
satellites. But the improvements gained by self-calibration are
significantly larger.

To see how the use of the $\chi$-$\lsat$ data in \mockb\ impacts the
purity and completeness, we also show the same statistics, but for
\mocka, in Figure \ref{f.comp2}. The comparisons between the standard
and self-calibrated algorithms are analogous to those in Figure
\ref{f.comp}. Differences in the overall curves at high luminosities,
$\lgal\gtrsim 10^{10}$ $\lsolhh$ are driven by the use of the
Blanton luminosity function, which falls off much more
rapidly than the Bernardi luminosity function.

\begin{figure*}
\plotone{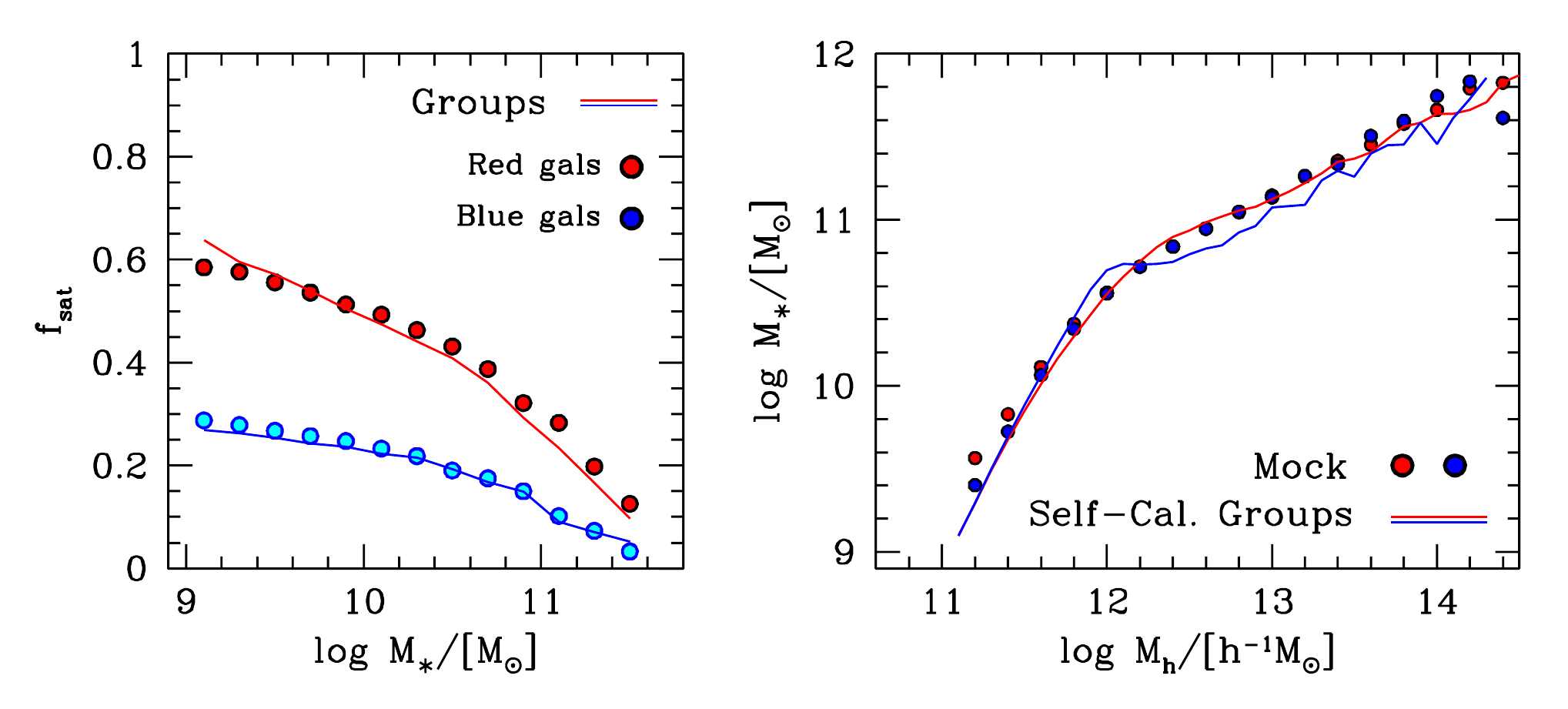}
\caption { {\it Left-Hand Panel:} Satellite fraction, $\fsat$, as a
  function of $\mgal$, for red and blue galaxies. Colored circles
  show values from \mockum. The
  solid curves show the predictions from the self-calibrated group
  finder. {\it Right-Hand Panel:} The SHMRs for red
  and blue galaxies. Filled circles show the values from \mockum. Solid
  curves show the predictions from the self-calibrated group
  finder. 
  \label{f.fsat_shmr_um}}
\end{figure*}

\subsection{UniverseMachine}
\label{s.results_um}

Figures \ref{f.wp_lsat_um} and \ref{f.fsat_shmr_um} show the mock
values from the UniverseMachine mock and the best-fit model of the
self-calibrating group finder. Here, stellar mass is the quantity by
which we rank-order the groups. Although the red galaxy clustering is
enhanced relative to the that of the blue galaxies, like the
luminosity mocks and the SDSS MGS results, $\fsat$ for red galaxies is
significantly lower than those results, while $\fsat$ for blue
galaxies is $\sim 50\%$ higher---$\fsat=0.3$ versus 0.2. The
$\lsatred/\lsatblue$ ratio is slightly above unity, even though the
SHMRs are approximately the same. The self-calibrated group finder
produces good fits to both the clustering and the $\lsat$ ratio, which
yield accurate predictions of the satellite fraction and the SHMRs for
the red and blue samples. The paucity of blue central galaxies in
halos above $10^{12}$ $\hmsol$ creates noise in the blue SHMR, but the
self-calibrated finder predicts that the SHMRs of the color
samples are roughly equal. The SHMR for blue galaxies is somewhat
underpredicted, by $\sim 0.15$ dex.

The gray shaded region shows the ratio total satellite {\it stellar}
mass, rather than our fiducial approach of satellite luminosity. This
is to compare our method of assigning $\lsat$ to UM halos---which is
based on $M_h$ only---to the intrinsic values from the UM model. As
discussed previously, galaxy quenching in UM depends on multiple
factors, including halo formation history, which may correlate with
the number of subhalos in a given halo. However, the $\lsat$ ratio is
consistent with the $M_{\ast, \rm sat}$ ratio, implying that the
empirically-driven quenching processes are not significantly impacting
the observables used in the self-calibration process.

\section{Caveats and Future Work}
\label{s.summary}

Although the self-calibrated group finding algorithm developed here
presents a potential solution to some of the problems inherent to group
finders, some issues remain.

\vspace{0.3cm}
\noindent {\it Group Centering:} This standard halo-based algorithm
constructs a group catalog where the central galaxy is always the
brightest in the halo. This is also the choice in the
percolation-based algorithm of \cite{berlind_etal:06_catalog}. The
algorithm of \cite{eke_etal:04} and \cite{robotham_etal:11} chooses
the brighter of the two galaxies closest to the luminosity-weighted
group center.  For the vast number of halos, these are reasonable
assumptions. However, given the scatter between $\mhalo$ and $\lgal$,
combined with the fact that $d\log\lgal/d\log\mhalo<1$ at high halo
masses, some fraction of halos will contain a piece of substructure
that contains a galaxy brighter than the central galaxy. This has been
detected, statistically, in observational data, by
\cite{skibba_etal:11}.

The self-calibrated algorithm we describe in the proceeding section
also makes this assumption. Improving on this assumption is a topic of
current research, which we leave to a future paper. We can estimate
the degree of the problem through our abundance matching calculations,
but the results depends on the assumptions made. Assuming the
\cite{blanton_etal:05} luminosity function and $\slogl=0.2$, for halos
with $\mhalo\ge 10^{13}$ $\hmsol$, a subhalo is brighter than the
central 26\% of the time. However, expressed as the fraction of groups
with centrals brighter than $\mr<-21$---the luminosity scale at which
the mean halo mass is also $\sim 10^{13}$ $\msol$, this occurs only
3.3\% of the time. These assumptions {\it maximize} this
frequency. \cite{lange_etal:19} constrain $\slogl$ using satellite
kinematics (as opposed to other recent works that measure $\slogm$) at
$\mhalo\gtrsim 10^{13}$ to be $\sim 0.15$, and to be a decreasing
function with increasing $\mhalo$. Using this value, and the
\cite{bernardi_etal:13} luminosity function, these values become 5.0\%
for halos and 1.5\% for groups.

As can be seen from our mock results, misidentification of the central
galaxy has limited impact on our results. The LHMR and SHMR values
recovered are unbiased relative to the input values in the mocks. The
most direct consequence is in estimating $\fsat$ for bright
galaxies---if a the brightest galaxy in a group is always considered
the central, this artificially suppresses $\fsat$ at the bright end
(\citealt{reddick_etal:13}). This effect can be seen in our results
(e.g., Figures \ref{f.fsat_shmr_mocka}, \ref{f.fsat_shmr_mockb}, and
\ref{f.fsat_shmr_um}).


\vspace{0.3cm}
\noindent {\it Galaxy Assembly Bias:} An inherent assumption in all
halo-based group-finders---indeed, in any group finder that uses
$\lgrp$ to estimate $\mhalo$---is that the amount of light in
satellite galaxies is a function of the halo mass only. For dark
matter halos, we know that the number of halos depends on the
formation history of the halo as well (\citealt{zentner_etal:05,
  gao_white:07, croft_etal:12, mao_etal:18}). At fixed $\mhalo$, older
halos have less substructure. This is one of the many manifestations
of halo assembly bias (e.g., \citealt{wechsler_tinker:18}). How much
this propagates into the population of satellite galaxies---e.g.,
whether there is significant galaxy assembly bias---is not
clear. \cite{alpaslan_tinker:19} find a small but non-zero trend of
$\lsat$ decreasing with large-scale density. The trend goes in the
direction expected---older halos live in higher-density environments,
thus the expectation is that $\lsat$ should decrease with
density. However, the amplitude of the change in $\lsat$ is only
$\sim 2\%$ at the highest and lowest values of density measured. If
galaxy assembly bias exists, it is not a strong effect on the total
luminosity in satellite galaxies.  Predictions of $\lsat$ from the
IllustrisTNG hydrodynamic galaxy formation simulation (\citealt{tng})
are in agreement with the observational results, showing little to no
correlation of $\lsat$ with large-scale density (Cao~et.~al., in
prepration).

An additional concern is the use of $\lsat$ data with secondary galaxy
properties. If a galaxy property correlates with halo formation
history at fixed $\mgal$, then the correlation of that property with
$\lgal$ would trace galaxy assembly bias, as opposed to constraining
$\mhalo$. The bimodal distribution of galaxy colors and star-formation
activity has been a focus of observational studies of assembly bias,
with most studies finding that the data are inconsistent with strong a
correlation between whether a galaxy is on the red sequence and halo
formation history, and weaker correlations are difficult to detect
observationally (\citealt{tinker_etal:08_voids, tinker_etal:17_p1,
  tinker_etal:18_p2, zu_mandelbaum:16, zu_mandelbaum:18,
  calderon_etal:18}). The lensing results of
\cite{mandelbaum_etal:16}, demonstrating that red central galaxies
live in more massive halos than blue centrals at fixed $\mgal$, are in
agreement with the $\lsat$ results for red and blue galaxy samples,
implying that $\lsat$ is probing $\mhalo$ for this galaxy property.

For the hypothetical galaxy property, $\chi$, an observational test
proposed by \cite{tinker_etal:19_lsat} is whether the value of $\chi$
correlates with large-scale density. \cite{alpaslan_tinker:19} find
that the properties of red central galaxies are all consistent with no
correlation with large-scale density, but for blue galaxies there are
non-zero correlations for stellar velocity dispersion, galaxy size,
and Sersic index. \cite{calderon_etal:18} also find a possible
assembly bias signal for Sersic index as well. This limits the choices
available for $\chi$ on real data, but candidates remain, including
galaxy concentration and surface density.

\vspace{0.3cm}
\noindent {\it Future Work:} The self-calibrated group finder is ready
to be run on SDSS MGS data. All clustering and $\lsat$ data are
already in place and results will come in a companion paper to this
one.

DESI-BGS has begun taking data, and by the end of the first year of
observations (out of five total) will have four times as many redshift
as the MGS. The longer redshift baseline afforded by the BGS, which
will target galaxies $\sim 2$ magnitudes deeper than the SDSS limit,
will open the door to investigating evolution of the galaxy-halo
connection with a single galaxy sample.

Future development of the self-calibration algorithm will also help
utilize the constraining power offered by ongoing and upcoming
surveys. Specifically in regards to the possibility of galaxy assembly
bias, rather than selecting galaxy properties that appear to lack any
detectable correlation with secondary halo properties, future
implementations may rather lean in to these observed signals, with the
goal of estimating not just the halo masses of galaxy groups, but
other halo properties as well. New data will be required to constrain
these new freedoms in the model, but upcoming data is well-positioned
to provide robust measurements of statistics sensitive to assembly
bias, including density correlations (\citealt{abbas_sheth:06,
  tinker_etal:18_p3}), marked correlation functions
(\citealt{skibba_etal:06, zu_mandelbaum:18, calderon_etal:18}), void
statistics (\citealt{tinker_etal:08_voids, walsh_tinker:19}), and the
cosmic web (\citealt{tojeiro_etal:17}).

\bibliography{risa}
\bibliographystyle{aasjournal}



\end{document}